\documentclass[10pt, twocolumns]{IEEEtran}
 \makeatletter

\newcommand{\Rmnum}[1]{\expandafter\@slowromancap\romannumeral #1@}
\makeatother
\usepackage{graphicx,cite,epsfig,amssymb,amsmath,multirow}
\usepackage{graphicx,amsmath,amssymb,amsfonts,cite}
\usepackage{mathrsfs}
\usepackage{algorithm}
\usepackage{algorithmic}
\usepackage{array}
\usepackage{float}
\usepackage{bm}
\usepackage{ulem}
\usepackage{color}

\usepackage{algorithm}
\usepackage{algorithmic}

\usepackage{amsthm}
\newtheorem*{theorem1}{Theorem 1 (Weyl)}
\newtheorem*{theorem2}{Theorem 2 (Wedin)}

\makeatother
\begin{document}
\bibliographystyle{ieeetr}

%------------------------------- title ----------------------------------------

\title{Smoothed SVD-based Beamforming for FBMC/OQAM Systems Based on Frequency Spreading}
%\author{\IEEEauthorblockN{Yu Qiu, Daiming Qu, Da Chen and Tao Jiang\\}
%\IEEEauthorblockA{School of Electronics Information and Communications\\
%		Huazhong University of Science and Technology, Wuhan, 430074, P. R. China\\
%		Email: qudaiming@hust.edu.cn}

	\author{\normalsize
	
	Yu Qiu, Daiming Qu, Da Chen, and Tao Jiang, \textit{Senior Member}, \textit{IEEE}
	\thanks{
		Yu Qiu, Daiming Qu (corresponding author), Da Chen, and Tao Jiang are with the School of Electronics Information and Communications, Huazhong University of Science and Technology, Wuhan, 430074, P. R. China (e-mail: qudaiming@hust.edu.cn).
		
		This work is supported in part by the National Natural Science Foundation of China (61571200, 61701186) and the open research fund of National Mobile Communications Research Laboratory, Southeast University (2014D09).
		
	}\\}

\markboth{} {{et al.}:Smoothed SVD-based Beamforming for FBMC/OQAM Systems Based on Frequency Spreading}

\maketitle
\begin{abstract}
The combination of singular value decomposition (SVD)-based beamforming and filter bank multicarrier with offset quadrature amplitude modulation (FBMC/OQAM) has not been successful to date. The difficulty of this combination is that, the beamformers may experience significant changes between adjacent subchannels, therefore destroy the orthogonality among FBMC/OQAM real-valued symbols, even under channels with moderate frequency selectivity. In this paper, we address this problem from two aspects: i) an SVD-FS-FBMC architecture is adopted to support beamforming with finer granularity in frequency domain, based on the frequency spreading FBMC (FS-FBMC) structure, i.e., beamforming on FS-FBMC tones rather than on subchannels; ii) criterion and methods are proposed to smooth the beamformers from tone to tone. The proposed finer beamforming and smoothing greatly improve the smoothness of beamformers, therefore effectively suppress the leaked ICI/ISI. Simulations are conducted under the scenario of IEEE 802.11n wireless LAN. Results show that the proposed SVD-FS-FBMC system shares close BER performance with its orthogonal frequency division multiplexing (OFDM) counterpart under the frequency selective channels.

\end{abstract}
%----------------------------- I. INTRODUCTION -----------------------------------
\section{Introduction}
Filter bank multicarrier with offset quadrature amplitude modulation (FBMC/OQAM) \cite{Farhang-Boroujeny2011, Siohan2002, Ihalainen2011, Gao2011, Hirosaki1981, Chen2013, Zhang2012, Wang2013,Cui2015,Qu2013tsp,Zhang2016,Lee2016} is considered as a promising alternative to the conventional orthogonal frequency division multiplexing (OFDM) technique \cite{Chang1968, Zou1995}. However, integration of multiple-input multiple-output (MIMO) techniques with FBMC/OQAM, in general, is more complicated than with OFDM. Thanks to the adding of CP, a subchannel (subcarrier band) is exactly flat and independent from other subchannels in OFDM systems. Thus, MIMO precoding and equalization can be taken on each subchannel independently, without leading to any inter-carrier interference (ICI) or inter-symbol interference (ISI). However, without CP, the MIMO precoding and equalization of FBMC systems are more complicated and could lead to considerable ICI/ISI under frequency selective channels, due to the fact that FBMC/OQAM is a non-orthogonal waveform (FBMC/OQAM symbols are orthogonal with each other only in the real domain \cite{Farhang-Boroujeny2011, Siohan2002, Hirosaki1981}) .

In FBMC/OQAM systems, the real and imaginary parts of QAM symbol are separated and transmitted as pulse amplitude modulated (PAM) symbols. There exists ICI/ISI interference between the PAM symbols in the form of imaginary interference. ICI/ISI-free symbols are obtained only after channel equalization and taking the real parts,
e.g., see \cite{Farhang-Boroujeny2011, Siohan2002, Hirosaki1981,Cui2015} for details. The combination of MIMO and FBMC/OQAM is a trivial task in channels with high coherence bandwidth, which is almost equivalent to MIMO-OFDM systems. While for the frequency selective channels, without carefully design, the beamforming matrices could differ dramatically between adjacent subchannels, and the imaginary interference from one subchannel could be leaked into adjacent subchannels as real interference, therefore destroy the orthogonality among FBMC/OQAM PAM symbols in the real domain \cite{Estella2010}.

Aware of the ICI/ISI interference, some studies \cite{Caus2014,Caus2013,Caus2012,Caus2015,Yao2014} attempt to constrain this interference by careful design of precoding as well as equalization for MIMO-FBMC/OQAM systems. In\cite{Caus2012}, two MIMO-FBMC precoding/equalization schemes were designed to maximize the signal to leakage plus noise ratio (SLNR) and the signal to interference plus noise ratio (SINR), respectively. Criterion of minimizing the sum mean square error was adopted in \cite{Caus2014}. The coordinated beamforming technique was applied in MIMO-FBMC/OQAM systems \cite{Yao2014}, where the precoding and decoding matrix are computed jointly and iteratively. A two-step method was proposed in \cite{Caus2015}, where the precoders are first optimized to maximize the SLNR given the equalizers and then, the equalizers are designed according to the minimum mean square error (MMSE) criterion while fixing the precoders. Although the ICI/ISI interference is suppressed, error performance loss or significantly increased complexity, compared with their OFDM counterparts, are observed with the aforementioned MIMO-FBMC/OQAM schemes. Very recently, a novel architecture was proposed to approximate an ideal frequency selective precoder
and linear receiver by Taylor expansion, exploiting the structure of the analysis and synthesis filter banks \cite{Mestre2016}. This architecture was shown to be very promising, however more results are needed to reveal its full potential. Smoothing of the precoders is proposed in \cite{Mestre2015}, which keeps the phase of one precoder component  constant accross subcarriers. However, the phase continuity crierion is not effective when this precoder component crosses zero and changes its sign. A more thorough review of MIMO-FBMC/OQAM precoding/beamforming techniques, including those for multi-user MIMO \cite{Candido2015,Newinger2014,Cheng2015}, could be found in \cite{Perez-Neira}.

%\textbf{WLP\cite{Cheng2013,MCaus2013}.
Most of the works on percoding/beamforming of MIMO-FBMC under frequency selective channels assume ployphase network implementation of FBMC \cite{Harris2003,Bellanger2010}. \color{black}In this paper, we focus on another type of implementation, namely frequency spreading FBMC (FS-FBMC), which has attracted wide attention in recent years \cite{Bellanger2012,Mattera2015,Dore2014,Mattera2015(1),Qu2016,Dore2017,Dore2015,Aminjavaheri2015,Won2017,Nadal2018,Mattera2017,Carvalho2017}. MIMO methods specifically designed for this type of implementation are in need. 

%, i.e., frequency spreading FBMC (FS-FBMC)  \cite{Bellanger2012,Mattera2015,Dore2014,Mattera2015(1),Qu2016}, are still in need. Please refer to  for recent research advances regarding the FS-FBMC implementation.

\color{black}Singular value decomposition (SVD) based beamforming, adopted in the IEEE 802.11n wireless LAN standard, yields maximum likelihood performance with simple linear transmit and receive beamformers for MIMO-OFDM systems \cite{Perahia2008}. Unfortunately, perfect combination with SVD beamforming is not yet available for FBMC/OQAM systems. In this paper, we propose a novel SVD-FS-FBMC scheme that is robust to channel frequency selectivity. While traditional beamforming of MIMO-FBMC are taken on per-subchannel basis, the proposed scheme enables beamforming of finer granularity in frequency by using the frequency spreading FBMC (FS-FBMC) structure \cite{Bellanger2012,Mattera2015,Dore2014,Mattera2015(1),Qu2016}, i.e., beamforming on FS-FBMC tones. We further propose two methods, namely phase factor optimization and orthogonal iteration, to smooth the SVD-based beamformers from tone to tone. The criterion of smoothing proposed is to minimize the Euclidean distance between adjacent beamformers. The proposed finer beamforming and smoothing methods greatly improve the smoothness of beamformers, therefore effectively suppress the leaked ICI/ISI from adjacent subchannels. Simulations are conducted under the scenario of IEEE 802.11n wireless LAN and the results show that the proposed SVD-FS-FBMC system performs closely with its OFDM counterpart under the IEEE 802.11n Channel Models. Our preliminary results on this subject have been reported in \cite{Qu2016}.

The following notations are used in this paper. Bold lower-case letters  denote column vectors. Bold upper-case letters are used for matrices. The superscripts $(\cdot)^{\textrm{T}}$, $(\cdot)^*$, $(\cdot)^{\textrm{H}}$, and $(\cdot)^\dag$ represent the transpose, conjugate, Hermitian transpose, and Moore-Penrose pseudo-inverse, respectively. $\Re\{\cdot\}$ and  $\Im\{\cdot\}$ denote the real and imaginary parts, respectively. $\mathbb{E}[\cdot]$ stands for  the expectation. $||\cdot||_2$ denotes vector or matrix 2-norm, $||\cdot||_F$ denotes the matrix Frobenius-norm, whenever the particular choice of norm is unimportant, $||\cdot||$ is used. \color{black} Function $\rm{sin}$ and $\rm{arccos}$ of a matrix is applied element-wisely in this paper. \color{black} Finally, ${\textrm j}=\sqrt{-1}$.

%All these works mentioned above try to cancel the interferece by complex precoding schemes. The performance is improved, but at a cost of great complexity.

%The authors in \cite{Caus2013} propose to concatenate two precoding matrices that one of the precoders is designed to cancel the interferences induced by the chanel and the remaining precoders are jointly designed with the receiver filters to minimizing the sum mean square error.

%In \cite{Caus2014}, two techniques are designed under the criterion of minimizing the sum mean square error. The first one is a suboptimal solution in resisting the effect of channel frequency selectivity in exchange for a reasonable complexity level, and the second option iteratively computes precoders and equalizers by resorting to an optimization performance, but at the expense of high complexity.

%which could mitigate co-channel interference by decomposing a MIMO channel into parallel and independent single-input single-output (SISO) subchannels.

\section{System Models \& ICI/ISI of SVD-FBMC/OQAM} \label{sec:basic FBMC/OQAM}
\subsection{FBMC/OQAM System Model}\label{subsec:FBMC/OQAM}
\begin{figure*}[!t]
	\centering
	\includegraphics[scale=0.6]{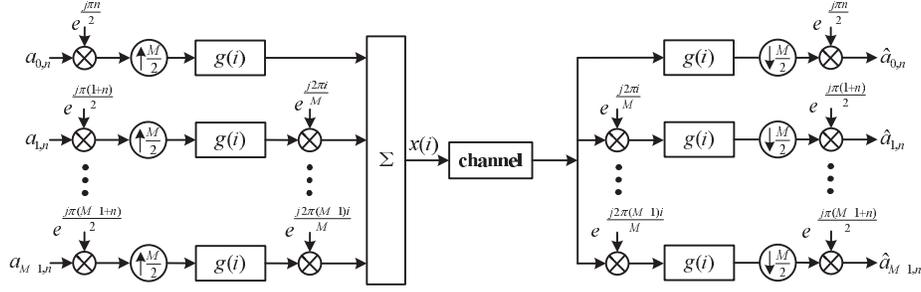}
	\caption{The equivalent baseband block diagram of an FBMC/OQAM system.}
	\label{fig:SystemModel}
\end{figure*}

Fig.~\ref{fig:SystemModel} presents the equivalent baseband block diagram of an FBMC/OQAM system. It consists of $M$ subcarriers with subcarrier spacing $1/T$,  where $T$ is the interval between the complex-valued symbols in time. Each complex-valued symbol is partitioned into a pair of real-valued PAM symbols. The PAM symbol at the frequency-time index $(m,n)$ is denoted by $a_{m,n}$, where $m$ is the frequency/subchannel index and $n$ is the time index. Moreover, $a_{m,2\hat{n}}$ and $a_{m,2\hat{n}+1}$, with integer $\hat{n}$, are  real and imaginary parts of a QAM symbol and are $T/2$ spaced in time. With a sampling interval of $T/M$, the filter bank prototype filter has the discrete time impulse response $g(i)$, which we assume only have non-zero coefficients for $0< i \le KM-1$, where $K$ is a positive integer. We further assume that $g(i)$ is an even-symmetric pulse, i.e., $g(i)=g(KM-i)$.

The discrete-time baseband equivalent of an FBMC/OQAM signal may be presented as \cite{2Lele2007,Siohan2002}
\begin{equation}\label{eq 1}
{x(i)}=\sum_{m=0}^{M-1}\sum_{n\in{\mathbb{Z}}}a_{m,n}\underbrace{g(i-n\frac{M}{2})e^{{\textrm j}\frac{2\pi mi}{M}}e^{{\textrm j}\frac{\pi(m+n)}{2}}}_{g_{m,n}(i)},
\end{equation}
where $g_{m,n}(i)$ is the pulse of the $(m,n)$-th PAM symbol. When an FBMC/OQAM signal is transmitted through a channel that varies slowly with time and its delay spread is significantly shorter than the symbol interval, the channel transfer function over each subchannel may be approximated by a flat gain. Let $H_{m,n}$ denote this gain for the $m$-th subchannel at the $n$-th time index. With the slow varying assumption, we omit the subscript $n$ from $H_{m,n}$ for simplicity of presentation. Then, the $m_0$th output of the receiver analysis filter bank at the $n_0$th time index is obtained as, \cite{2Lele2007,Siohan2002},
\begin{equation}\label{eq 2}
\begin{split}
r_{m_0,n_0} &= \sum_{i=-\infty}^{\infty}g_{m_0,n_0}^{*}(i)\sum_{m=0}^{M-1}\sum_{n \in {\mathbb{Z}}}H_{m}a_{m,n}g_{m,n}(i) \\
&\approx H_{m_0}a_{m_0,n_0} + H_{m_0}\sum _{(m,n)\neq (m_0,n_0)}a_{m,n}{\zeta _{m,n}^{m_0,n_0}}
\end{split}
\end{equation}
where
\begin{equation}\label{eq 5}
{\zeta _{m,n}^{m_0,n_0}}=
\sum_{i=-\infty}^{\infty}g_{m_0,n_0}^{*}(i)g_{m,n}(i).
\end{equation}

It is noteworthy that for a well-designed prototype filter $g(i)$, ${\zeta _{m,n}^{m_0,n_0}=1}$ when $(m,n)=(m_0,n_0)$, and is zero or a pure imaginary value when $(m,n)\neq (m_0,n_0)$, e.g., see \cite{Farhang-Boroujeny2011}. To be more specific, ${\zeta _{m,n}^{m_0,n_0}}$, for $(m,n)\neq (m_0,n_0)$, represents the imaginary interference to $a_{m,n}$, which could be removed by taking the real part after channel equalization.

\subsection{SVD-based MIMO Beamforming}\label{subsec:SVD}
We consider a MIMO system equipped with ${N_\textrm{t}}$ transmit antennas and ${N_\textrm{r}}$ receive antennas, which supports ${L}$ parallel streams. The SVD decomposition of the channel ${\bf{H}} \in {\mathbb{C}^{{N_\textrm{r}} \times {N_\textrm{t}}}}$ is:
\begin{equation}\label{eq 4.1}
{\bf{H}} = {\bf{U}}{\bf{D}}{\bf{V}}^{\textrm H},
\end{equation}
where ${\bf{V}}\in {\mathbb{C}^{{N_\textrm{t}}\times{N_\textrm{t}}}}$ and ${\bf{U}}\in {\mathbb{C}^{{N_\textrm{r}}\times{N_\textrm{r}}}}$ are unitary matrices, and ${\bf{D}}$ is an $N_\textrm{r}$-by-$N_\textrm{t}$ rectangular diagonal matrix containing $(\lambda^1,\lambda^2\ldots)$ as diagonal elements, where $\lambda^1,\lambda^2\ldots$ denote the singular values that are sorted in descending order and are real-valued. When ${L=N_\textrm{t}=N_\textrm{r}}$, the transmit beamformer and receive beamformer are simply ${\bf{V}}$ and ${\bf{U}}^{\textrm H}$, respectively. When $L<N_\textrm{t}$ or $L<N_\textrm{r}$, they are submatrices of $\bf{V}$ and ${\bf{U}}^{\textrm H}$, respectively, corresponding to the $L$ largest singular values. For clarity of notations, we abuse the notations $\bf{V}$ and/or ${\bf{U}}^{\textrm H}$ and let them also represent the beamformers when $L<N_\textrm{t}$ or $L<N_\textrm{r}$ in the rest of this paper, then ${\bf{V}}\in {\mathbb{C}^{{N_\textrm{t}}\times{L}}}$ and ${\bf{U}}^{\textrm H}\in {\mathbb{C}^{{L}\times{N_\textrm{r}}}}$. With the transmit beamforming and MIMO channel, the signal at the receiver is
\begin{equation}\label{eq 4.1}
\bf{y} = \bf{H}\bf{V}\bf{s} + \bf{n},
\end{equation}
where $\bf{s}$ is the symbols to be transmitted and $\bf{n}$ is the receive noise vector ($\bf{y}$, $\bf{s}$ and $\bf{n}$ are ${N_\textrm{r}}\times{1}$, ${L}\times{1}$ and ${N_\textrm{r}}\times{1}$ vectors, respectively). The signal after receive beamforming is
\begin{equation}\label{eq 4.1}
{\bf{r}}= {\bf{U}}^{\textrm H}{\bf{y}}
= {\bf{U}}^{\textrm H}({\bf{H}}{\bf{V}}{\bf{s}} + {\bf{n}})
= {\bf{D}}{\bf{s}} + \widetilde{\bf{n}},
\end{equation}
%\begin{equation}
 %\widetilde{\bf{a}}_{{m_0},{n_0}}={\bf{a}}_{{m_0},{n_0}}+\bf{I}+\wid%etilde{\bf{n}}.
%\end{equation}
where $\widetilde{\bf{n}}={\bf{U}}^{\textrm H}{\bf{n}}$ ($\bf{r}$ and $\widetilde{\bf{n}}$ are ${L}\times{1}$ vectors). Clearly, the transmitted symbols are recovered with nonequal gains \cite{Lebrun2005}, and there is no interference among streams.

\subsection{Straightforward Combination of SVD and FBMC/OQAM}\label{subsec_straight}
In this subsection, we present the model of straightforward combination of SVD and FBMC/OQAM and discuss about the interference leakage problem. Combining the models in subsection \ref{subsec:FBMC/OQAM} and \ref{subsec:SVD}, the transmitted signal of the SVD-FBMC/OQAM is represented by a sequence of vectors ${\bf{x}}(i) = [{x_1}(i)~{x_2}(i)\cdot\cdot\cdot{x_{N_\textrm{t}}}(i)]^{\textrm{T}}$ as
\begin{equation}\label{eq 4.1}
{\bf{x}}(i) = \sum\limits_{m = 0}^{M-1} {\sum\limits_{n \in {\mathbb{Z}}} { {{{\bf{V}}_{m}}{\bf{a}}_{m,n}{g_{m,n}}(i)} } },
\end{equation}
where ${\bf{a}}_{m,n} = [a_{m,n}^{1}~a_{m,n}^{2}\cdot\cdot\cdot a_{m,n}^{L}]^{\textrm{T}}$ is the symbol vector to be transmitted, and ${g_{m,n}}(i)$ is the frequency-time shifted version of prototype filter $g(i)$ (see (\ref{eq 1})), ${{\bf{V}}_{m}} \in {\mathbb{C}^{{N_\textrm{t}} \times {L}}}$ denotes the beamforming matrix for the $m$-th subchannel. Assuming nearly flat fading across the subcarrier band (the bandwidth of subchannel), the received signal at the output of analysis filters for Time $n_0$ and Subchannel $m_0$ is
\begin{equation}\label{eq 2.f}
\begin{split}
{{\bf{r}}_{{m_0},{n_0}}} \approx& \sum\limits_{m = 0}^{M - 1} {\sum\limits_{n \in {\mathbb{Z}}} {{\bf{U}}_{{m_0}}^{\textrm H}{{\bf{H}}_m}} {{\bf{V}}_m}{{\bf{a}}_{m,n}}\sum\limits_{i =  - \infty }^\infty  {{g_{{m_0},{n_0}}^*(i)g_{m,n}}(i)} } \\
=& \sum\limits_{m = 0}^{M - 1} {\sum\limits_{n \in {\mathbb{Z}}}{{\bf{U}}_{{m_0}}^{\textrm H}{{\bf{H}}_m}{{\bf{V}}_m}{{\bf{a}}_{m,n}}} {\zeta _{m,n}^{{m_0},{n_0}}}} \\
=&~{{\bf{D}} _{m_0}}{{\bf{a}}_{{m_0},{n_0}}}   \\
&+\sum\limits_{(m,n) \ne ({m_0},{n_0})} {{\bf{U}}_{{m_{\bf{0}}}}^{\textrm H}{{\bf{H}}_{m}}{{\bf{V}}_m}{{\bf{a}}_{m,n}}} {\zeta _{m,n}^{{m_0},{n_0}}},\\
\end{split}
\end{equation}
where ${{\bf{H}}_{m}} \in {\mathbb{C}^{{N_\textrm{r}} \times {N_\textrm{t}}}}$ is the MIMO channel response of Subchannel $m$, ${\bf{U}}_{m_0}^{\textrm H}\in {\mathbb{C}^{{L} \times {N_\textrm{r}}}}$ is the receive beamformer for the $m_0$-th subchannel. Clearly, the first term of (\ref{eq 2.f}) is the recovered symbols and the second term is the ICI/ISI interference. If ${\bf{H}}_{{m_0}}\approx{\bf{H}}_{{m}}$ and ${\bf{V}}_{{m_0}}\approx{\bf{V}}_{{m}}$ for subchannels adjacent to $m_0$, we have ${{\bf{U}}_{{m_0}}^{\textrm H}{{\bf{H}}_{m}}{{\bf{V}}_m}} \approx {{\bf{D}} _{m_0}}$, and the ISI/ICI term is approximately imaginary and could be removed by taking the real part of ${{\bf{r}}_{{m_0},{n_0}}}$ (recall that ${{\bf{D}} _{m_0}}$ is real-valued). However, this does not work even under channels with moderate frequency selectivity. The reason is that: the transmit and receive beamformer may experience significant changes between adjacent subchannels due to channel variation, then ${{\bf{U}}_{{m_0}}^{\textrm H}{{\bf{H}}_{m}}{{\bf{V}}_m}}$ is not real-valued and the ISI/ICI term is no longer pure imaginary, which results in leaked ICI/ISI interference into the real part of ${{\bf{r}}_{{m_0},{n_0}}}$.

\section{The Finer Beamforming Architecture}
In this section, we propose finer beamforming for FBMC/OQAM, here finer beamforming means beamforming with finer granularity in frequency domain. As discussed in the above section, the straightforward SVD-FBMC/OQAM systems beamform at the subchannel level, i.e., each subchannel has its own transmit and receive beamformer. To enable a finer granularity, we adopt the FS-FBMC structure \cite{Bellanger2012,Mattera2015,Dore2014,Mattera2015(1)}, and beamformers are designed at each tone of FS-FBMC. The proposed finer beamforming provides a basic architecture to support smoother changes from subchannel to subchannel.

\subsection{Frequency Spreading FBMC (FS-FBMC)}
The FS-FBMC structure, a special form of the fast convolution implementation of filter banks \cite{Boucheret1999,Zhang2000,Pucker2003,Umehira2010,Renfors2011,Renfors2014}, uses frequency spreading/despreading to implement the filtering in the frequency domain for FBMC/OQAM systems. In FS-FBMC, FFT/IFFT is taken at the length of $KM$. Let $G(k)$ denote the FFT of a segment of $g(i)$ in the range of $0\leq i\leq{KM-1}$, where $0\leq k\leq{KM-1}$ is the index of FS-FBMC tones. We assume that $G(k)$ has $2P-1$ non-negligible tones that center around the zero-th tone, where $P$ is a positive integer. Let $G_{m,n}^{(n_0)}(k)$ denote the FFT of a segment of $g_{m,n}(i)$ in the range of $n_0M/2\leq i\leq{n_0M/2+KM-1}$, which is the portion of $g_{m,n}(i)$ that falls inside the $n_0$-th sliding window. The superscript $(n_0)$ is to emphasize that the FFT is taken at the $n_0$-th window. Then, the filtering at the $n$-th time index is implemented in the frequency domain by spreading the PAM symbols with $G_{m,n}^{(n)}(k)$ as
\begin{equation}\label{eq 3.1.1}
b_{n}(k) =\sum\limits_{m = 0}^{M - 1} {  a_{m,n}{G_{m,n}^{(n)}(k)}},
\end{equation}
where $G_{m,n}^{(n)}(k)$ is the FFT of the non-zero part of pulse $g_{m,n}(i)$, i.e., at the $n$-th window. And, the output of the frequency spreading is fed to the IFFT transformation to obtain the time domain samples of the $n$-th time index as
\begin{equation}\label{eq 3.2}
x_{n}(i) =
\begin{cases}
\begin{split}
&\sum\limits_{k = 0}^{KM - 1} b_n(k) {e^{{\textrm j}\frac{{2\pi k(i-nM/2)}}{{KM}}}},\\
& ~~~~~~~~\frac{nM}{2} \leq i \leq \frac{nM}{2}+KM-1
\end{split}
\\
0,~~~~~~{\rm else}
\end{cases}.
\end{equation}
Then, the transmitted time sequence is obtained by accumulation over all symbols as
\begin{equation}\label{eq 3.2}
x(i) = \sum\limits_{n \in {\mathbb{Z}}} {x_{n}(i )}.
\end{equation}

At the receiver, a sliding window is employed to select $KM$ samples every $M/2$ samples, which are fed to an FFT module. Then, the transmitted PAM symbols are recovered through equalization and frequency despreading. More details of FS-FBMC transmission could be found in \cite{Berg2014}.

\subsection{Finer SVD Beamforming} \label{sec:finer svd beamforming}
\begin{figure}[!h]
	\centering
	\includegraphics[scale=0.63]{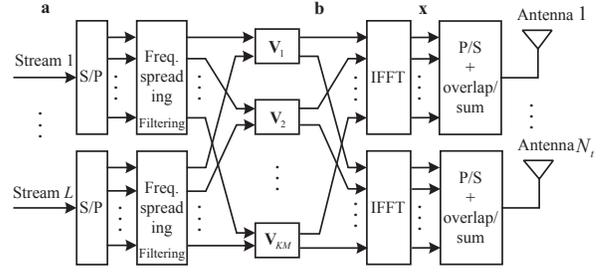}
	\caption{The transmitter of the proposed SVD-FS-FBMC scheme.}
	\label{fig:transmiter}
\end{figure}

Fig.~\ref{fig:transmiter} presents the transmitter of the proposed finer SVD-FS-FBMC, where ${\bf{V}}_k \in {\mathbb{C}^{{N_\textrm{t}} \times {L}}}$ is the beamforming matrix for the $k$-th FS-FBMC tone, and ${\bf{V}}_k = [{\bf{v}}_k^1~{\bf{v}}_k^2 \cdot\cdot\cdot {\bf{v}}_k^{L}]$, here ${\bf{v}}_k^l$ denotes the beamforming vector for the $l$-th stream. The beamformed signals on all transmit antennas for the $k$-th tone and $n$-th time index is given by an $N_\textrm{t}$-by-1 vector
\begin{equation}\label{eq 3.1}
{{\bf{b}}_n}(k) = {{\bf{V}}_k}\sum\limits_{m = 0}^{M - 1} {{{\bf{a}}_{m,n}{G_{m,n}^{(n)}(k)}} }.
\end{equation}
Then, the samples on all transmit antennas for the $n$-th time index is obtained by IFFT
\begin{equation}\label{eq def_x}
{\bf{x}}_{n}(i) =
\begin{cases}
\begin{split}
&\sum\limits_{k = 0}^{KM - 1}{{\bf{b}}_n}(k){e^{{\textrm j}\frac{{2\pi k(i-nM/2)}}{{KM}}}}, \\
&~~~~~~~~\frac{nM}{2} \leq i \leq \frac{nM}{2}+KM-1
\end{split}
\\
0,~~~~~~{\rm else}
\end{cases},
\end{equation}
where ${\bf{x}}_{n}(i)$ is an $N_\textrm{t}$-by-$1$ vector.

After accumulation over all symbols, the transmitted signals on all antennas are represented by the following vector sequence
\begin{equation}\label{eq 3.1}
{\bf{x}}(i) = \sum\limits_{n \in {\mathbb{Z}}} {{{\bf{x}}_n}(i } ).
\end{equation}

\begin{figure}[!h]
	\centering
	\includegraphics[scale=0.6]{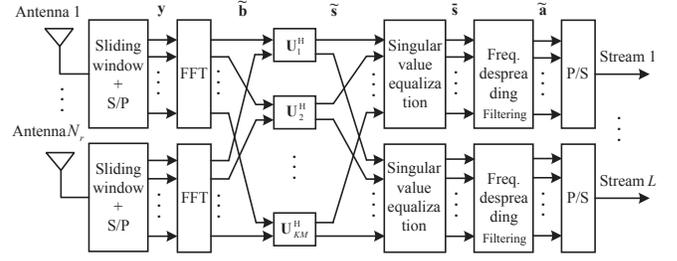}
	\caption{The receiver of the proposed SVD-FS-FBMC scheme.}
	\label{fig:receiver}
\end{figure}

Fig.~\ref{fig:receiver} presents the receiver of the proposed SVD-FS-FBMC, where ${\bf{U}}^{\textrm H}_k \in {\mathbb{C}^{{L} \times {N_\textrm{r}}}}$ is the beamforming matrix for the $k$-th FS-FBMC tone, and ${\bf{U}}_k^{\textrm H} = [({\bf{u}}_k^1)^{\textrm H}~({\bf{u}}_k^2)^{\textrm H} \cdot\cdot\cdot ({\bf{u}}_k^{L})^{\textrm H}]$, here $({\bf{u}}_k^l)^{\textrm H}$ denotes the receive beamforming vector for the $l$-th stream.
Let $N_\textrm{r}$-by-1 vector ${\bf{y}}(i)$ denote the $i$-th received samples on all receive antennas, and ${\bf{y}}^{(n_0)}(i)$ ($0\leq i\leq{KM-1}$) denote the selected $KM$ samples by the $n_0$-th sliding window, i.e., ${\bf{y}}^{(n_0)}(i)={\bf{y}}(i+n_0M/2)$, for $0 \le i \le KM-1$. Taking FFT of ${\bf{y}}^{(n_0)}(i)$, we obtain
\begin{equation}\label{eq 3.B.1}
\begin{split}
\widetilde{{\bf{b}}}^{(n_0)}(k)=\frac{1}{KM}{\sum\limits_{i = 0}^{KM - 1} {{\bf{y}}^{(n_0)}(i){e^{ - {\textrm j}\frac{{2\pi ki}}{{KM}}}}} }, 0\leq k\leq{KM-1}.
\end{split}
\end{equation}

Applying the receive beamformer, we have the signals received on the $k$-th tone
\begin{equation}\label{eq 3.B.2}
\widetilde{{\bf{s}}}^{(n_0)}(k)={\bf{U}}^{\textrm H}_k\widetilde{{\bf{b}}}^{(n_0)}(k),
\end{equation}
where $\widetilde{{\bf{s}}}^{(n_0)}(k)$ is an $L$-by-$1$ vector that holds the samples of all streams. Noting that different tones and streams may have different singular values, an equalizer is required at each tone for each stream. The equalizers for the $k$-th tone is represented by an $L$-by-$L$ diagonal matrix ${\bf{E}}_k={\rm{diag}}\left \{E_k^1,E_k^2,\ldots, E_k^{L}\right\}$, where $E_k^l$ is the equalizer for Stream $l$ and Tone $k$. If it is a zero forcing (ZF) equalization, $E_k^l=1/\lambda_k^l$. Then, the equalized signal is obtained as
\begin{equation}
\overline{{\bf{s}}}^{(n_0)}(k)={\bf{E}}_k\widetilde{{\bf{s}}}^{(n_0)}(k).
\end{equation}
Finally, despreading is applied to obtain the PAM symbols at the $m_0$-th subchannel and $n_0$-th time index
\begin{equation}\label{eq final_pre}
\begin{split}
{\widetilde{\bf{a}}_{m_0,n_0}} = & \sum\limits_{k = 0}^{KM - 1}G_{m_0,n_0}^{(n_0)\,*}(k)\overline{{\bf{s}}}^{(n_0)}(k)\\
= & \sum\limits_{k = 0}^{KM - 1}G_{m_0,n_0}^{(n_0)\,*}(k){{\bf{E}}_k}{{\bf{U}}_k^{\textrm H}}\widetilde{{\bf{b}}}^{(n_0)}(k).
\end{split}
\end{equation}

Under the assumption that ${\bf{E}}_k$, ${\bf{U}}_k$, ${\bf{H}}_k$ and ${\bf{V}}_k$ are nearly flat, i.e., they vary slowly, across the subcarrier band,
\begin{equation}\label{eq final}
\begin{split}
&{\widetilde{\bf{a}}_{m_0,n_0}} \\
\approx~&{\bf{a}}_{m_0,n_0} +\sum\limits_{(m,n) \ne (m_0,n_0)}{{\bf{a}}_{m,n} {\sum\limits_{k = 0}^{KM - 1} {\zeta _{m,n}^{m_0,n_0}}} } \\
&+{ {\sum\limits_{k = 0}^{KM - 1} {G_{m_0,n_0}^{(n_0)\,*}(k){{\bf{E}}_k}{{\bf{U}}_k^{\textrm H}}{\bf{n}}^{(n_0)}(k)} } },
\end{split}
\end{equation}
where ${\bf{n}}^{(n_0)}(k)$ is FFT of the noise with ${\bf{y}}^{(n_0)}(i)$. Proof of this equation is presented in the Appendix. Clearly, the first term of (\ref{eq final}) is the transmitted PAM symbols, and the second term is the ICI/ISI interference that is pure imaginary under the nearly-flat assumption and could be removed by taking the real part.

Obviously, the complexity of beamforming in the proposed SVD-FS-FBMC system is roughly proportional to the number of tones per subchannel, i.e. $K$. When $K=1$, i.e. without frequency spreading and despreading, the proposed SVD-FS-FBMC is exactly the same as that of SVD-OFDM with the same subchannel number $M$. Therefore, the complexity of beamforming in the proposed SVD-FS-FBMC is $K$ times that of the SVD-OFDM with the same subchannel number $M$.

Compared with the straightforward SVD-FBMC/OQAM given in Section \ref{subsec_straight}, the major improvement by the finer beamforming is that it allows smoother transition of beamformers.

%Using (\ref{eq 3.B.1}) and (\ref{eq 3.B.2}), the estimated PAM symbols at Time $n_0$ and Subchannel $m_0$ is
%\begin{equation}\label{eq 3.1}
%\begin{split}
%{\widetilde{\bf{a}}_{m_0,n_0}} \approx   &{ {\sum\limits_{k = 0}^{KM - 1} {G_{m_0,n_0}^{(n_0)\,*}(k){{\bf{E}}_k}{{\bf{U}}_k^{\textrm H}}\times }}}\\ &{{{({{\bf{H}}_k}{{\bf{V}}_k}\sum\limits_{m = 0}^{M - 1}\sum\limits_{n \in {\mathbb{Z}}} {{{\bf{a}}_{m,n}{G_{m,n}^{(n_0)}(k)}} }  + {\bf{n}}^{(n_0)}(k) )}}}.
%\end{split}
%\end{equation}
%Using the relation that ${{\bf{U}}_k^{\textrm H}}{{\bf{H}}_k}{{\bf{V}}_k}= {{\bf{D}}_k}$, ${\bf{E}}_k{\bf{D}}_k$ $={\bf{I}}_L$ (assuming the ZF equalization of singular values) and $\sum_{k = 0}^{KM - 1} G_{m_0,n_0}^{(n_0)\,*}(k) {{G_{m,n}^{(n_0)}(k)}}$ $=\sum_{i=-\infty}^{\infty}g_{m_0,n_0}^{*}(i)g_{m,n}(i)$ $={\zeta _{m,n}^{m_0,n_0}}$, we have
%\begin{equation}\label{eq 3.21}
%\begin{split}
%&{\widetilde{\bf{a}}_{m_0,n_0}} \\
%\approx~&  {\bf{a}}_{m_0,n_0}{\sum\limits_{k = 0}^{KM - 1} {G_{m_0,n_0}^{(n_0)\,*}(k) {{{G_{m_0,n_0}^{(n_0)}(k)}} } } } \\
%& +\sum\limits_{(m,n) \ne (m_0,n_0)}{{\bf{a}}_{m,n} {\sum\limits_{k = 0}^{KM - 1} {G_{m_0,n_0}^{(n_0)\,*}(k) {{{G_{m,n}^{(n_0)}(k)}} } } } } \\
%&+{ {\sum\limits_{k = 0}^{KM - 1} {G_{m_0,n_0}^{(n_0)\,*}(k){{\bf{E}}_k}{{\bf{U}}_k^{\textrm H}}{\bf{n}}_{n_0}(k)} } }\\
%\approx~&{\bf{a}}_{m_0,n_0} +\sum\limits_{(m,n) \ne (m_0,n_0)}{{\bf{a}}_{m,n} {\sum\limits_{k = 0}^{KM - 1} {\zeta _{m,n}^{m_0,n_0}}} } \\
%&+{ {\sum\limits_{k = 0}^{KM - 1} {G_{m_0,n_0}^{(n_0)\,*}(k){{\bf{E}}_k}{{\bf{U}}_k^{\textrm H}}{\bf{n}}_{n_0}(k)} } }.
%\end{split}
%\end{equation}

\section{Bounds on the Euclidean Distance between Adjacent Beamforming Matrices}\label{sec analysis flat}
The derivation of the finer SVD-FS-FBMC in Subsection \ref{sec:finer svd beamforming} is under the assumption that ${\bf{H}}_k$ is nearly flat across the subcarrier bandwidth, as well as ${\bf{E}}_k$, ${\bf{U}}_k$ and ${\bf{V}}_k$. In this section, we will show that ${\bf{E}}_k$, ${\bf{U}}_k$ and ${\bf{V}}_k$ that are nearly flat are available, i.e., they can be bounded in Euclidean distance between adjacent tones, as long as ${\bf{H}}_k$ is nearly flat. The reasoning here is based on the perturbation theory for SVD decomposition \cite{perturbation,H.Weyl1912,Wedin1972}.

This section also serves as a justification to the smoothing criterion proposed in Section \ref{sec smooth}, which minimizes the Euclidean distance between beamformers of adjacent tones.

As the channel is assumed to be nearly flat, ${\bf{H}}_k$ can be written as
\begin{equation}
{\bf{H}}_{k}={\bf{H}}_{k-1}+\Delta{\bf{H}}_{k},
\end{equation}
where $\Delta{\bf{H}}_{k}$ is an $N_r$-by-$N_t$ matrix, and $||\Delta{\bf{H}}_{k}||_2 \ll ||{\bf{H}}_{k-1}||_2$. Then, the Weyl theorem \cite{H.Weyl1912} gives a bound on the difference between the singular values of ${\bf{H}}_{k}$ and ${\bf{H}}_{k-1}$.
\begin{theorem1}\label{Theorem 1}
	\begin{equation}
 |{\lambda_{k}^l}-\lambda_{k-1}^l| \le ||\Delta{\bf{H}}_{k}||_2,~~~l=1,\cdot\cdot\cdot,L.
 \end{equation}
\end{theorem1}
Due to the Weyl theorem, when $||\Delta{\bf{H}}_{k}||_2$ is small, the difference between $\lambda_{k}^l$ and $\lambda_{k-1}^l$ is also small. Thus, ${\bf{E}}_k$ can be assumed nearly flat.

%The singular vectors can also be bounded.

To bound the difference between ${\bf{V}}_k$ and ${\bf{V}}_{k-1}$ as well as that between ${\bf{U}}_k$ and ${\bf{U}}_{k-1}$, we take use of the Wedin theorem \cite{Wedin1972} below. With the Wedin's theorem, we will show that the eigenspaces spanned by ${\bf{V}}^l_k$ and ${\bf{V}}^l_{k-1}$ are close, where ${\bf{V}}^l_k=[{\bf{v}}_k^1,\cdots,{\bf{v}}_k^l]$. We will also show that ${\bf{V}}^{l-1}_k$ and ${\bf{V}}^{l-1}_{k-1}$ are close. Then, we reach the conclusion that the eigenspaces spanned by ${\bf{v}}_k^l$ and ${\bf{v}}_{k-1}^l$ are close. Similarly, subspaces ${\bf{u}}_k^l$ and ${\bf{u}}_{k-1}^l$ are close as well.

In stead of directly bounding the Euclidean distance between the singular vectors, the Wedin theorem gives a bound on angles between the subspaces spanned by the singular vectors. \color{black} Let ${\bf{L}}$ and ${\bf{M}}$ in $\mathbb{C}^{N \times l}$ have full column rank $l$, the angle matrix from ${\bf{L}}$ to ${\bf{M}}$ is defined as \cite{Wedin1972}
%\begin{equation}\label{eq_angle_define}
%||\sin {\bf{\Theta}}({\bf{L}},{\bf{M}})||_F=  ||({\bf{I}}-{\bf P}_{{\bf{M}}}){\bf P}_{{\bf{L}}}||_F,
%\end{equation}
%where ${\bf{\Theta}}({\bf{L}},{\bf{M}})$ is an $l$-by-$l$ real matrix, which is defined as
\begin{equation} 
	\begin{split}
&{\bf{\Theta}}({\bf{L}},{\bf{M}})=\\
&~~{\rm{arccos}}{(({\bf{L}}^{\textrm H}{\bf{L}})^{-\frac{1}{2}}}{\bf{L}}^{\textrm H}{\bf{M}}({\bf{M}}^{\textrm H}{\bf{M}})^{-1}{\bf{M}}^{\textrm H}{\bf{L}}({\bf{L}}^{\textrm H}{\bf{L}})^{-\frac{1}{2}})^{-\frac{1}{2}}. \nonumber
	\end{split}
\end{equation}
%${\bf P}_{\bf{L}}$ and ${\bf P}_{\bf{M}}$ are the othogonal projections onto the subspace ${\bf{L}}$ and ${\bf{M}}$, respectively, and $||\cdot||_F$ denotes the matrix Frobenius-norm. 
Then, $||\sin {\bf{\Theta}}({\bf{L}},{\bf{M}}) ||_F$ gives a measure of how much the subspaces of ${\bf{L}}$ and ${\bf{M}}$ are separated in angle \cite{Wedin1972}. With the definition, the Wedin theorem is given as
\begin{theorem2}\label{Theorem 2} If there is a $\delta>0$, such that
	\begin{equation}\label{wedin 1}
	\min_{1 \le i \le l, j \ge l+1} |\lambda_{k}^i-\lambda_{k-1}^j| \ge \delta,
	\end{equation}
    and
    \begin{equation}\label{wedin 2}
    \lambda_k^l \ge \delta,
    \end{equation}
    then
   \begin{align}\label{equation wedin}  
   &\sqrt {||\sin {\bf{\Theta}}({\bf{V}}_k^l,{\bf{V}}_{k-1}^l) ||_F^2 + ||\sin {\bf{\Theta}}({\bf{U}}_k^l,{\bf{U}}_{k-1}^l) ||_F^2} \nonumber \\
   &\le \frac{{\sqrt {||{\bf{R}}_{\rm R}^l||_F^2 + ||{\bf{R}}_{\rm L}^l||_F^2} }}{\delta },
   \end{align}
where
%\begin{equation}
%\begin{split}
%{\bf{R}}_l=&~{\bf{H}}_k{\bf{v}}_{k}^l-{\bf{u}}_{k}^l\lambda_k^l \\
%{\bf{S}}_l=&~{\bf{H}}^{\rm {H}}_k %{\bf{u}}_{k}^l-{\bf{v}}_{k}^l\lambda_k^l.
%\end{split}
%\end{equation}
\begin{equation}
	\begin{split}
		{\bf{R}}_{\rm R}^l=&~{\bf{H}}_{k-1}{\bf{V}}_{k}^l-{\bf{U}}_{k}^l{{\rm{Diag}}(\lambda_k^1,\cdots,\lambda_k^l)} \\
		{\bf{R}}_{\rm L}^l=&~{\bf{H}}^{\rm {H}}_{k-1} {\bf{U}}_{k}^l-{\bf{V}}_{k}^l{{\rm{Diag}}(\lambda_k^1,\cdots,\lambda_k^l)}.
	\end{split}
\end{equation}
\end{theorem2}
\color{black} The bound (\ref{equation wedin}) is a combined bound. The left-hand side combines the angles for the left and right singular subspace. The right-hand side combines what might be called right and left residuals. The conditions (\ref{wedin 1}) and (\ref{wedin 2}) are separation conditions. The first says that $\lambda_{k}^1,\cdots, \lambda_{k}^l$ are separated from $\lambda_{k-1}^{l+1},\cdots$. The second condition says that the singular value $\lambda_{k}^l$ are separated from the ghost singular values (singular values very close to zero).

In addition to the Wedin theorem, the following inequalities hold \cite{perturbation}
\begin{equation}\label{eq_RlSl}
\begin{split}
||{\bf{R}}_{\rm R}^l||\le  ||\Delta{\bf{H}}_{k}||,
||{\bf{R}}_{\rm L}^l||\le  ||\Delta{\bf{H}}_{k}||.
\end{split}
\end{equation}

Combining the Wedin theorem and (\ref{eq_RlSl}), it is clear that, subspaces ${\bf{U}}_k^l$ and ${\bf{V}}_k^l$ are stable, i.e., the angles between ${\bf{U}}_k^l$ and ${\bf{U}}_{k-1}^l$, plus angles between ${\bf{V}}_k^l$ and ${\bf{V}}_{k-1}^l$, are bounded by (\ref{equation wedin}), under the conditions (\ref{wedin 1}) and (\ref{wedin 2}). In other words, subspaces ${\bf{V}}_k^l$ and ${\bf{V}}_{k-1}^l$ \big(subspaces ${\bf{U}}_k^l$ and ${\bf{U}}_{k-1}^l$\big) are close to each other.

Similarly, if the seperation conditions are satisfied for $l-1$, subspaces ${\bf{V}}_k^{l-1}$ and ${\bf{V}}_{k-1}^{l-1}$ are close to each other. Since ${\bf{V}}_k^{l}=[{\bf{V}}_k^{l-1} ~ {\bf{v}}_k^{l}]$ and ${\bf{V}}_{k-1}^{l}=[{\bf{V}}_{k-1}^{l-1} ~ {\bf{v}}_{k-1}^{l}]$, we can say that ${\bf{v}}_k^l$ and ${\bf{v}}_{k-1}^l$ are close, measured by angles between subspaces. Similarly, subspaces ${\bf{u}}_k^l$ and ${\bf{u}}_{k-1}^l$ are close as well. The bounded angles between subspaces means that, given any ${\bf{v}}_{k-1}^l$ (or ${\bf{u}}_{k-1}^l$), a vector with bounded Euclidean distance to ${\bf{v}}_{k-1}^l$ (or ${\bf{u}}_{k-1}^l$) is available in subspace ${\bf{v}}_k^l$ (or ${\bf{u}}_k^l$).

% ${\bf{v}}_{k}^l$ (or ${\bf{u}}_{k}^l$)

Concluding the discussion above, when ${\bf{H}}_k$ is assumed nearly flat, ${\bf{U}}_k$ and ${\bf{V}}_k$ that are nearly flat are available, i.e., beamformers of adjacent tones can be bounded in Euclidean distance, if all corresponding singular values satisfy the separation conditions (\ref{wedin 1}) and (\ref{wedin 2}).

%provides bound for both right and left singular subspaces

%Clearly, the canonical angles for the left and right singular subspace are both bounded. The distance of the subspace between  ${\bf{v}}_{k}^1$(${\bf{u}}_{k}^1$) and ${\bf{v}}_{k-1}^1$(${\bf{u}}_{k-1}^1$) is bounded.  Therefore, the singular vectors of a matrix are also perfectly conditioned- the perturbation on the singular vectors are bounded within the norm of the perturbation. A vector can be picked from $\Pi({\bf{v}}_k^1)$($\Pi({\bf{u}}_k^1)$), which has the bounded Euclidean distance with ${\bf{v}}_{k-1}^1$(${\bf{u}}_{k-1}^1$). The analysis also applicable to other left and right singular vectors ${\bf{u}}_{k}^i$ and ${\bf{v}}_{k}^i$ ($i=2,\cdot\cdot\cdot L$). ${\bf{V}}_k$ and ${\bf{U}}_k$ can be regarded as nearly flat when ${\bf{H}}_k$ is nearly flat.

\section{Smoothing of Beamformers}\label{sec smooth}
In Section \ref{sec analysis flat}, the theories show that ${\bf{E}}_k$, ${\bf{U}}_k$ and ${\bf{V}}_k$ can be bounded in Euclidean distance between adjacent tones, as long as ${\bf{H}}_k$ is nearly flat and the separation conditions are satisfied for singular values of all streams. However, unfortunately, not every SVD algorithm generates ${\bf{U}}_k$ and ${\bf{V}}_k$'s that are bounded in Euclidean distance across adjacent $k$'s as required by the proposed architecture. The reason is that: the SVD decomposition is not unique, it may produce more than one set of singular vectors that span the same space but are different \cite{Sandell2009}.

%in this section The proposed finer beamforming architecture itself does not guarantee a low leaked ICI/ISI, it is valid only when the beamformers are smooth across the transmit bandwidth (see Appendix). Therefore, the SVD decomposition to be used in this architecture should be able to provide smooth transition from ${\bf{V}}_{k-1}$ to ${\bf{V}}_k$ for all $k$s. In this section, a direct optimization method and an orthogonal iteration method are presented.

\subsection{Smoothing Criterion and Phase Factor Optimization}

To deal with this problem, we propose in this subsection a smoothing criterion and method to smooth the output of any given SVD algorithm so that it satisfies the nearly-flat requirement. The idea here is as follows: We smooth the output of the given SVD algorithm, denoted by ${\hat{\bf{V}}}_k$, to obtain ${\bf{V}}_k$ such that its Euclidean distance to ${\bf{V}}_{k-1}$ is bounded as given in Section \ref{sec analysis flat}. The criterion of minimizing the distance of adjacent beamformers is intuitive, due to the fact that Euclidean distance, i.e., Euclidean difference of adjacent vectors, is a measure of first-order smoothness for vector functions. In addition, the criterion is known to be effective thanks to the discussion in Section \ref{sec analysis flat}. Running this operation from the beginning to the end of the active frequency tones, a sequence of smoothed beamformers is obtained. It is worth mentioning here that we have also attempted to optimize the second-order smoothness, however no further observable gain over the first-order smoothness optimization was obtained. The reason is that the channel frequency response is rather smooth with the finer beamforming and second-order smoothness or above is unnecessary, under the system parameters and channel models considered in Section \ref{simulation}.

It is assumed that ${\hat{\bf{v}}_{k}^l}$ is the $l$-th right singular vector given by the SVD algorithm, corresponding to the singular value ${\hat{\lambda}_k^l}$ of ${{\bf{H}}_k}$, here the hat $\hat{~}$ is to emphasize that ${\hat{\bf{v}}_{k}^l}$ and ${\hat{\lambda}_k^l}$ are the outputs of the SVD algorithm before smoothing. Taking ${\hat{\bf{v}}_{k}^l}$ as the input, after smoothing,  ${{\bf{v}}_{k}^l}$ is output as the beamformer.

The subspace spanned by ${\hat{\bf{v}}_{k}^l}$ can be represented by a set ${\bf \Pi}(\hat{{\bf{v}}}_k^l)=\{\beta{\hat{\bf{v}}_{k}^l}{e^{j\theta }}|\beta \in \mathbb{R}, \theta \in [0,2\pi)\}$. If ${\hat{\lambda}_k^l}$ satisfy the separation conditions, distance from ${\bf \Pi}(\hat{{\bf{v}}}_k^l)$ to ${\bf \Pi}({{\bf{v}}}_{k-1}^l)$ is bounded due to the Wedin theorem, here ${{\bf{v}}}_{k-1}^l$ is the smoothed beamformer of the $(k-1)$-th frequency tone. Therefore, a vector that has a bounded Euclidean distance to ${{\bf{v}}}_{k-1}^l$ is available in ${\bf \Pi}(\hat{{\bf{v}}}_k^l)$. Then, the problem of smoothing the SVD output is solved by
\begin{equation}\label{eq 5_v1}
{\bf{v}}_k^l=e^{j\theta^* }{\hat{\bf{v}}_{k}^l},
\end{equation}
where ${e^{j\theta^* }}$ is a phase factor that minimizes the distance from ${e^{j\theta }}{\hat{\bf{v}}_{k}^l}$ to ${\bf{v}}_{k-1}^l$, i.e., 
\begin{equation}\label{eq 4.1}
\begin{split}
\theta^*=\mathop {\textrm{arg\,min}}\limits_\theta &\{||{e^{j\theta }}{\hat{\bf{v}}_{k}^l} - {\bf{v}}_{k-1}^l||_2\}.
\end{split}
\end{equation}

The solution to the phase factor optimization problem is
\begin{equation}\label{eq 5_phase}
{e^{j\theta^* }} = \frac{{(\hat{\bf{v}}_{k}^l)^{\textrm{H}}{\bf{v}}_{k-1}^l}}{{|(\hat{\bf{v}}_{k}^l)^\textrm{H}{\bf{v}}_{k-1}^l|}}.
\end{equation}
Combining (\ref{eq 5_v1}) and (\ref{eq 5_phase}), we have the smoothed output as
\begin{equation}\label{eq 4.1}
{\bf{v}}_k^l=\frac{{(\hat{\bf{v}}_{k}^l)^\textrm{H}{\bf{v}}_{k-1}^l}}{{|(\hat{\bf{v}}_{k}^l)^\textrm{H}{\bf{v}}_{k-1}^l|}}{\hat{\bf{v}}_{k}^l}.
\end{equation}

%The beamforming vectors for each tone is designed according the former tone. Hence, the difference between adjacent beamforming vectors is minimized. The beamforming vector for other tones and streams are analyzed as above.

% ${\hat{\bf{v}}_{k}^l}$ is also an eigenvector of ${{\bf{A}}_k}={{\bf{H}}_k^\textrm{H}}{{\bf{H}}_k}$.

%Also, the beamforming vectors are the columns of unitary matrix, therefore $\beta=1$, and the beamforming vector set ${\varPsi}_k^l$ for the $k$-th tone and $l$-th stream can be presented as ${\hat{\bf{v}}_{k}^l}{e^{j\theta }}$. Clearly, there is a phase factor ${e^{j\theta }}$  difference between any two vectors in set ${\varPsi}_k^l$.

%The beamforming vector set $\mathcal{C}$ for the $k$-th tone and $l$-th stream can be presented as $e^{j\theta}\hat{\bf{v}}_{k}^l$.
% In addition, $\textrm{rank}({\bf{A}}_k-\lambda_k^l {\bf{I}})=N_t-1$.
% \begin{equation}
% {{\bf{A}}_k}(\beta {\hat{\bf{v}}_{k}^l}{e^{j\theta }}) ={\lambda_k^l}(\beta {\hat{\bf{v}}_{k,1}^l}{e^{j\theta }}).
% \end{equation}

%To enable the smoothness between adjacent beamforming vectors, a vector ${\bf{v}}_{k}^l$, which has the  minimum Euclidean distance with ${\bf{v}}_{k-1}^l$, will be picked from set ${\varPsi}_k^l$, where ${\bf{v}}_{k-1}^l$ is the expected $(k-1)$-th tone and $l$-th stream beamforming matrix. ${\bf{v}}_k^l$ is determined as below:

When two singular values become very close to each other, the bound given by the Wedin theorem is not close to zero even when the channel frequency response is smooth, then smoothness between adjacent beamformers cannot be guaranteed. Still, smoothing is needed to minimize the Euclidian distance of ${\bf{v}}_{k}^l$ and ${\bf{v}}_{k-1}^l$ for all $l$'s.
While smoothing the SVD output in this case, a special problem needs to be treated carefully. The problem is the ambiguity in pairing one from $\{\hat{\bf{v}}_k^{l},\ldots,\hat{\bf{v}}_k^{L}\}$ with ${\bf{v}}_{k-1}^l$, which is explained in the following. When $\hat{\lambda}_{k}^l \approx \lambda_{k-1}^l$ and it is separated from other singular values, there is no doubt that $\hat{\bf{v}}_k^{l}$ should be paired with ${\bf{v}}_{k-1}^{l}$. However, when there are multiple singular values of the $k$-th frequency tone approximate $\lambda_{k-1}^l$, i.e., $\hat{\lambda}_{k}^{l_1} \approx \ldots \approx \hat{\lambda}_{k}^{l_m} \approx \lambda_{k-1}^l$, where $m$ is the number of singular values that are close to $\lambda_{k-1}^l$,  there has to be a method to determine which of $\hat{\lambda}_{k}^{l_1},\ldots,\hat{\lambda}_{k}^{l_m}$ ($\hat{\bf{v}}_k^{l_1},\ldots,\hat{\bf{v}}_k^{l_m}$) should be paired with $\lambda_{k-1}^l$ (${\bf{v}}_{k-1}^{l}$), i.e., which one belongs to the $l$-th stream.

To resolve this ambiguity, we measure the subspace distance from $\hat{{\bf{v}}}_k^{l_i}$ to ${{\bf{v}}}_{k-1}^l$, for ${l_1,\ldots,l_m}$, and select the one with the minimum subspace distance to pair with ${{\bf{v}}}_{k-1}^l$. The subspace distance of two vectors, ${\hat{\bf{v}}_k^{{l_i}}}$ and ${\hat{\bf{v}}_{k-1}^{{l}}}$ in our discussion, is defined as \cite{Golub1996}
%\begin{equation}
%d({\bf{X}},{\bf{Y}})\buildrel \Delta \over = ||{\bf{X}}{\bf{X}}^\textrm{H}-{\bf{Y}}{\bf{Y}}^\textrm{H}||_2.
%\end{equation}
%Then, we have
\begin{equation}
d_{l_i}=||{\hat{\bf{v}}_k^{{l_i}}}({\hat{\bf{v}}_k^{{l_i}}})^{\textrm{H}}-{{\bf{v}}_{k-1}^l}({{\bf{v}}_{k-1}^l})^{\textrm{H}}||_2,
\end{equation}
and let
\begin{equation}
l_*=\mathop {\textrm{arg\,min}}\limits_{l_i \in {l_1,\ldots,l_m}}\{d_{l_i}\}.
\end{equation}
Then, $\hat{\lambda}_k^{l_*}$  is taken as the singular value of Stream $l$, and ${{\hat{\bf{v}}}_k^{{l_*}}}$ is smoothed to generate the beamformer as
\begin{equation}
\begin{split}
\lambda_k^l=&\hat{\lambda}_k^{l_*}\\
{\bf{v}}_k^l=&\frac{{({{\hat{\bf{v}}}_k^{{l_*}}})^\textrm{H}{\bf{v}}_{k-1}^l}}{{|({{\hat{\bf{v}}}_k^{{l_*}}})^\textrm{H}{\bf{v}}_{k-1}^l|}}{{{\hat{\bf{v}}}_k^{{l_*}}}}.
\end{split}
\end{equation}

%As a final remark of this subsection, it is noted that the proposed phase factor optimization is different from the phase continuity design proposed by \cite{Mestre2015} in that, phase is not the optimization objective but used as variable in this paper to ensure variation of eigenvectors follow a smooth path. Instead of phase continuity, the first-order smoothness, i.e., the Euclidean distance between adjacent eigenvectors is employed as the optimization objective.

%When the channel eigenvalues become close to each other, there exist ambiguity in matching ${\bf{v}}_{k-1}^l (l=1,\cdot\cdot\cdot,L)$ to $\hat{\bf{v}}_k^{l'} (l'=1,\cdot\cdot\cdot,L)$ \cite{Mestre2015}. To minimize the distance between adjacent beamforming vector, we need to find a vector $\hat{\bf{v}}_k^{l'}$ with the minimize subspace distance  with ${\bf{v}}_{k-1}^l$. If ${\bf{X}}$ and ${\bf{Y}}$ are matrices with orthonormal columns, the distance between the subspaces they span can be defined as \cite{Golub1996}

%The ${\hat{\bf{v}}_k^l}$ with the minimum Euclidean distance with ${\bf{v}}_{k-1}^l$ is the beamforming vector correspond to the $l$-th stream. Then,  ${\hat{\bf{v}}_k^l}$ is updated by direct optimization method.

%The phase ambiguity may occurs in the situation that the channel eigenvalues become close to each another, which can also be settled by the subspace determination. And then the direct phase optimization is adopted to get the optimal beamforming matrices for each tone.

\subsection{Smoothing by Orthogonal Iteration}
%The proposed finer beamforming architecture itself does not guarantee a low leaked ICI/ISI, it is valid only when the beamformers are smooth across the transmit bandwidth. Therefore, the SVD decomposition to be used in this architecture should be able to provide smooth transition from ${\bf{V}}_{k-1}$ to ${\bf{V}}_k$ for all $k$s. To satisfy this requirement,
In this subsection, we introduce the orthogonal iteration method \cite{Golub1996} for smoothing, which spontaneously follows the proposed smoothing criterion due to its iterative nature. Compared with the phase factor optimization method, the orthogonal iteration enjoys lower computational complexity. The orthogonal iteration method has been used in \cite{Sandell2009} to provide smooth beamforming for an OFDM system to enable channel state information (CSI) smoothing.

As we know, ${\bf{V}}_k$ is the right singular vectors of ${\bf{H}}_k$, as well as the eigenvectors of ${\bf{A}}_k = {\bf{H}}_k^{\textrm H}{\bf{H}}_k$, which can be found by performing the following iteration from an initial matrix
${{\bf{Q}}^{(0)}}\in{\mathbb{C}^{{N_\textrm{t}} \times {L}}}$ with orthonormal columns
\begin{equation}\label{eq 4.1}
\begin{split}
&{{\bf{B}}^{(i)}} = {{\bf{A}}_k}{{\bf{Q}}^{(i - 1)}},~~i = 1,2,\cdot\cdot\cdot  \\
&{\rm{QR~decomposition:~}}{{\bf{B}}^{(i)}} = {{\bf{Q}}^{(i)}}{{\bf{R}}^{(i)}} ,
\end{split}
\end{equation}
where $i$ denotes the iteration index and $N_{\rm{iter}}$ is the total number of iterations. According to \cite{Sandell2009}, ${{\bf{R}}^{(i)}}$ converges to a diagonal matrix containing the eigenvalues of ${{\bf{A}}_k}$, and ${{\bf{Q}}^{(i)}}$ converges to an orthonormal basis for the dominant subspace of dimension $L$.

For a beamforming that is smooth from Tone $k-1$ to $k$, ${\bf{V}}_{k-1}$ could serve as the initial ${{\bf{Q}}^{(0)}}$, and the output is ${\bf{V}}_{k}={{\bf{Q}}^{(N_{\rm{iter}})}}$. It will be shown in the next section that very few iterations are needed to obtain a satisfactory ${\bf{V}}_{k}$, which ensures smoothness from ${\bf{V}}_{k-1}$ to ${\bf{V}}_{k}$. The complete algorithm is presented in Algorithm 1.

\renewcommand{\thealgorithm}{1:}
\begin{algorithm}[htb]
	%\begin{algorithm}[tb]
	\caption{Orthogonal Iteration for Smooth Beamforming}
	\begin{algorithmic}[1]
		\label{alg:RPM}
		\vspace{7pt}
		\STATE Initialize ${\bf{V}}_0=\rm{SVD}({\bf{H}}_0)$~~~($\rm{SVD}(\cdot)$ stands for an arbitrary SVD algorithm);
		\FOR {$k = 1: KM-1$}
		\STATE ${\bf{A}}_k={\bf{H}}_k^{\textrm H}{\bf{H}}_k$;
		\STATE ${\bf{V}}_k={\bf{V}}_{k-1}$;
		\FOR {$i = 1: N_{\rm{iter}}$}
		\STATE ${\bf{B}}_k={\bf{A}}_k {{\bf{V}}_k}$;
		\STATE Update ${\bf{V}}_k$ using the following QR~decomposition: ${\bf{B}}_k={\bf{V}}_k {{\bf{R}}_k}$;
		\ENDFOR
		\STATE ${\bf{D}}_k=\rm{SQRT}({\bf{R}}_k)$ ($\rm{SQRT}(\cdot)$ stands for the square root of an diagonal matrix);
		\ENDFOR    			
	\end{algorithmic}
\end{algorithm}

\section{Simulation Results}\label{simulation}
In this section, we evaluate the performance of the finer and smoothed SVD beamforming for FBMC/OQAM that was proposed in this paper, through computer simulations. We compare the proposed SVD-FS-FBMC system with an SVD-OFDM system, under a setup similar to the IEEE 802.11n wireless LAN standard. Thanks to orthogonality among subchannels, the error performance of SVD-OFDM is the upper bound of the proposed SVD-FS-FBMC, if the CP overhead of OFDM is ignored. The presented results reveal excellent performance of our proposed method, which can compete with OFDM and give close BER results with 64-QAM constellation, under channel models of the IEEE 802.11n standard. \color{black} As shown by Table \ref{tab_channel}, the Channel Model D, E, and F of the IEEE 802.11n standard have relatively large maximum delay spread, when normalized by the OFDM symbol duration ($3200$ns), which result in strong frequency selectivity. Especially for the Channel Model F, the maximum delay spread is even longer than the CP defined in the standard ($800$ns).

\begin{table}[ht]	

	\centering 
\color{black}	\caption{Maximum Delay Spread of the Channel Models}
\label{tab_channel}
\begin{tabular}{|m{1.2cm}<{\centering\arraybackslash }|m{2.5cm}<{\centering\arraybackslash}|m{3cm}<{\centering\arraybackslash }|}
			\hline
		Channel Model & Maximum delay spread (ns) & Maximum delay spread normalized by the OFDM symbol duration \\
		\hline
		D & 390 & 12.2\% \\
		\hline
		E & 730 & 22.8\% \\
		\hline
		F & 1050 & 32.8\% \\
		\hline
		
	\end{tabular}

\end{table}

\color{black} For the simulations presented in this section, the following parameters are used for both SVD-FS-FBMC and SVD-OFDM systems. The MIMO system is configured as $N_\textrm{t}=N_\textrm{r}=L=2$. There are $M=64$ subcarriers, and the subcarrier spacing is $312.5$~kHz. Among the $64$ subcarriers, $48$ are active subcarriers modulated with 16-QAM or 64-QAM constellations. We apply no power allocation among subcarriers and streams in the simulation, i.e., all streams (and $48$ subcarriers) have equal transmit power. For channel coding, we use convolutional code of rate $2/3$ and constraint length $7$. A random interleaver is applied after the coding. Each data frame consists of $7$ FBMC/OFDM symbols (each consists of $48$ OQAM/QAM symbols). The FBMC systems employ the PHYDYAS filter \cite{Phydyas}, and the overlapping factor $K$ is $4$. With the FFT size of $4M$, the filter has $7$ non-negligible tones, i.e., $P=4$. For the singular value equalization, a ZF equalizer is employed at receiver of the proposed SVD-FS-FBMC. And, we set $N_{\rm{iter}}=3$ for the orthogonal iteration method. The SNR in the simulation is defined as: ${\rm{SNR}} \stackrel{\Delta}{=}{{N_\textrm{t}}\sigma _a^2\sigma _h^2}/{\sigma _n^2}$, where $\sigma _a^2$, $\sigma _h^2$ and $\sigma _n^2$ are the expected signal power of each transmit antenna, expected channel power gain between a pair of transmit and receive antennas, and expected AWGN noise power of each receive antenna, respectively, on each active subchannel.

In most of the following figures, we present the performance of the proposed SVD-FS-FBMC system with the orthogonal iteration of three iterations. To justify the use of orthogonal iteration and $N_{\rm{iter}}=3$, BER performance comparison between the orthogonal iteration of different iterations and phase factor optimization is presented in Section \ref{simulation_complexity}, followed by complexity comparison of these proposed smoothing methods.

\begin{figure}[!h]
	\centering
	\includegraphics[scale=0.6]{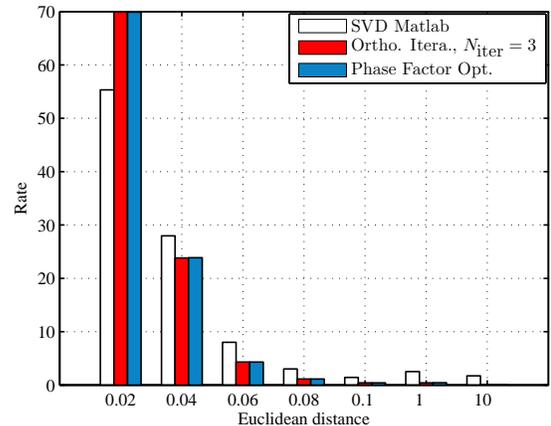}
	\caption{The Euclidean distance between transmit beamformers of adjacent tones.}
	\label{fig:distance}
\end{figure}

\subsection{Smoothness of Beamformers}
Let us first check if the smoothness across tones is improved with the proposed smoothing methods introduced in Section \ref{sec smooth}. Smoothness between two beamformers of adjacent tones is measured by their Euclidean difference. Fig.~\ref{fig:distance} presents the histogram of the Euclidean distance between adjacent beamformers by the proposed methods. Result by the SVD function in the Matlab software is also presented for comparison. The distance of the proposed schemes falls in the range of $0 \sim 1.0$, while that of SVD Matlab could go beyond $1.0$ with non-negligible percentage. It is thus concluded that the phase factor optimization and the orthogonal iteration method provide significant smoothness improvements compared with the SVD of no smoothness consideration. It is also observed that there is no observable difference in the histogram between the phase factor optimization and orthogonal iteration with $N_{\rm{iter}}=3$, which verifies the ability of the orthogonal iteration to fulfill the proposed smoothing criterion spontaneously.

%\begin{figure}[!h]
%	\centering
%	\includegraphics[scale=0.6]{BER_QPSK.eps}
%	\caption{BER performance of the SVD-FS-FBMC systems with QPSK and rate 2/3 coding, subchannel level or finer beamforming, orthogonal iteration or Matlab SVD.}
%	\label{fig:BER_QPSK}
%\end{figure}

\subsection{The BER Performance}

\begin{figure}[!h]
	\centering
	\includegraphics[scale=0.6]{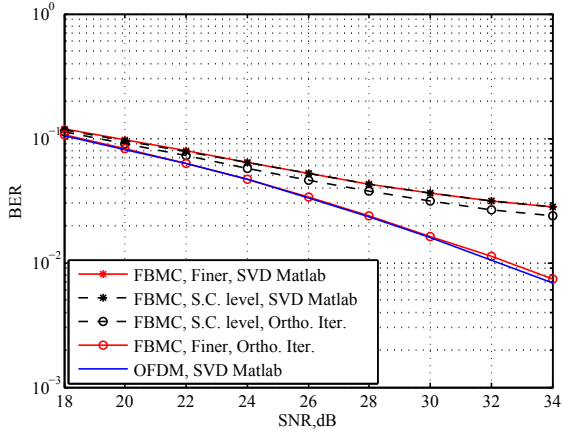}
	\caption{BER performance of the SVD-FS-FBMC systems with 64-QAM and no coding, subchannel level or finer beamforming, orthogonal iteration or Matlab SVD, Channel Model D, $N_{\rm{iter}}=3$.}
	\label{fig:BER_64QAM_NoCoding}
\end{figure}

\begin{figure}[!h]
	\centering
	\includegraphics[scale=0.6]{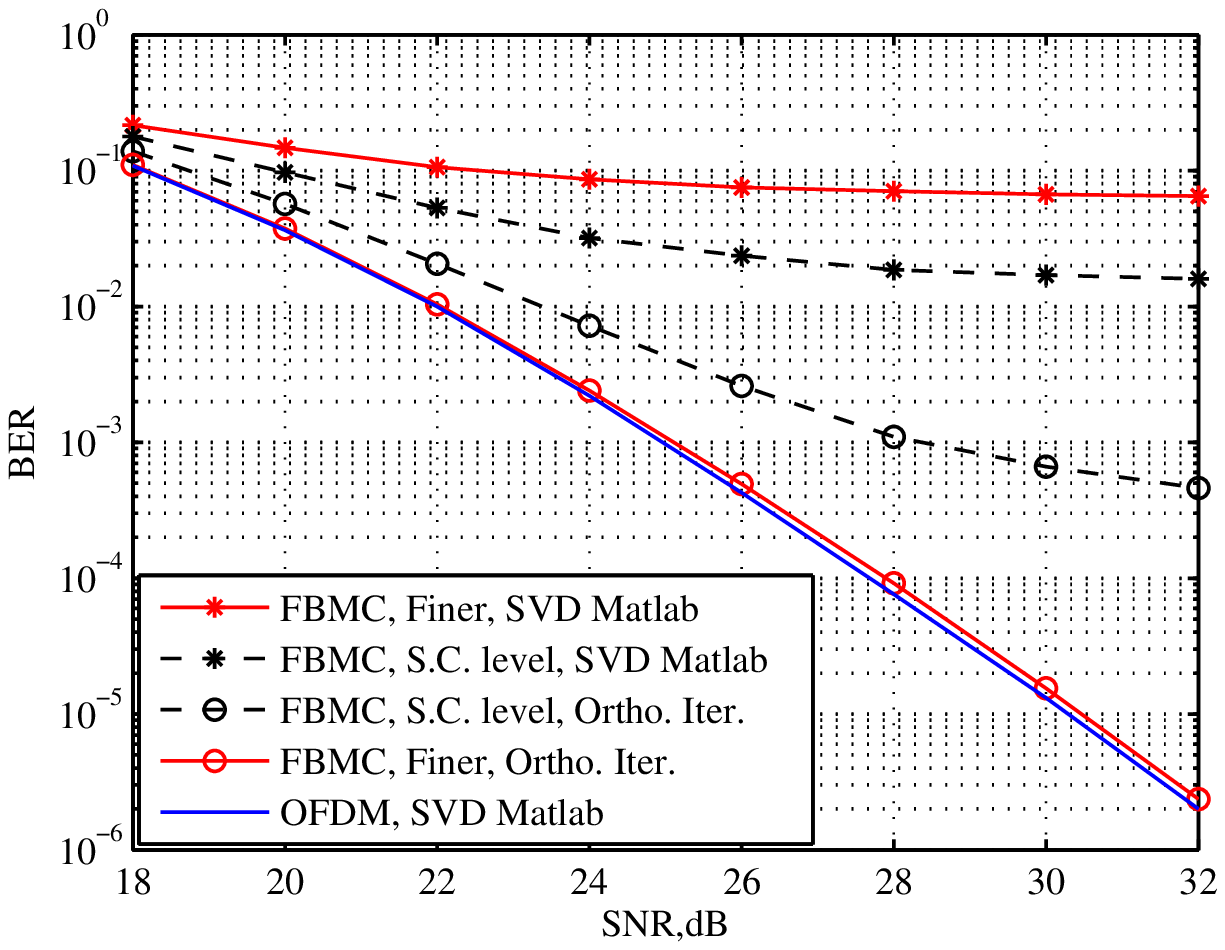}
	\caption{BER performance of the SVD-FS-FBMC systems with 64-QAM and rate 2/3 coding, subchannel level or finer beamforming, orthogonal iteration or Matlab SVD, Channel Model D, $N_{\rm{iter}}=3$.}
	\label{fig:BER_64QAM}
\end{figure}

\begin{table}[ht]
	\centering 
	\color{black}	\caption{Summary of the Schemes under Comparison}	
	\label{tab_comp}	
	\begin{tabular}{|m{3.0cm}<{\centering\arraybackslash }|m{1.7cm}<{\centering\arraybackslash }|m{1.9cm}<{\centering\arraybackslash }|}
		
		\hline
		Schemes&	Beamforming granularity	&Smoothing\\

		\hline
		SVD-OFDM	&S.C. level	&SVD Matlab (No smoothing)\\
		\hline
		Basic SVD-FBMC/OQAM	&S.C. level	&SVD Matlab (No smoothing)\\
		\hline
		SVD-FBMC/OQAM w/ smoothing	&S.C. level	&Ortho. Iter. (Smoothing)\\
		\hline
		SVD-FS-FBMC w/o smoothing	&Finer	&SVD Matlab (No smoothing)\\
		\hline
		Proposed SVD-FS-FBMC	&Finer&	Ortho. Iter. (Smoothing)\\
		\hline
		
	\end{tabular}
	
\end{table}

\color{black}The simulation results in this subsection mainly demonstrate the performance of the orthogonal iteration method, the comparison between the orthogonal iteration method and the phase factor optimization method is presented in Subsection \ref{simulation_complexity}.
Fig. \ref{fig:BER_64QAM_NoCoding} and \ref{fig:BER_64QAM} present BER performance of the SVD-FS-FBMC system with the proposed finer beamforming and smoothing under Channel Model D, without and with coding, respectively. \color{black}For comparison, we also present BER results of the following four systems: i) SVD-OFDM; ii) Basic SVD-FBMC/OQAM with subchannel-level (S.C. level) beamforming and without smoothing (SVD Matlab), this is the straightforward combination of SVD and FBMC/OQAM discussed in Section \ref{sec:basic FBMC/OQAM}; iii) SVD-FBMC/OQAM with subchannel-level beamforming and smoothing (orthogonal iteration), its performance gap to the proposed SVD-FS-FBMC shows how the proposed beamforming with finer granularity improves the performance; iv) SVD-FS-FBMC with the finer beamforming but without smoothing (SVD Matlab), its performance gap to the proposed SVD-FS-FBMC shows how the proposed smoothing improves the performance. \color{black} The schemes under comparison are summarized in Table \ref{tab_comp}. The simulations of Fig. \ref{fig:BER_64QAM_NoCoding} and \ref{fig:BER_64QAM} demonstrate that both beamforming with finer granularity and smoothing are necessary for a good beamforming of FBMC system under frequency selective channels. \color{black}The results clearly show that the SVD-FS-FBMC system with the proposed finer beamforming and smoothing greatly outperforms the other SVD-FBMC/OQAM and SVD-FS-FBMC systems. And it performs very closely with SVD-OFDM, in terms of BER, under the IEEE 802.11n Channel Model D. It is also observed that smoothing is crucial for the BER performance: the subchannel-level SVD-FBMC/OQAM systems with smoothing outperforms the SVD-FBMC/OQAM without smoothing. One may notice that the system with finer beamforming but no smoothing has the worst BER performance among the systems in comparison. The reason is that, when finer beamforming is employed without smoothing, significant changes of beamformer may happen between two tones of one subcarrier band, which results in serious distortion of the transmitted signal.

\begin{figure}
	\centering
	\includegraphics[scale=0.6]{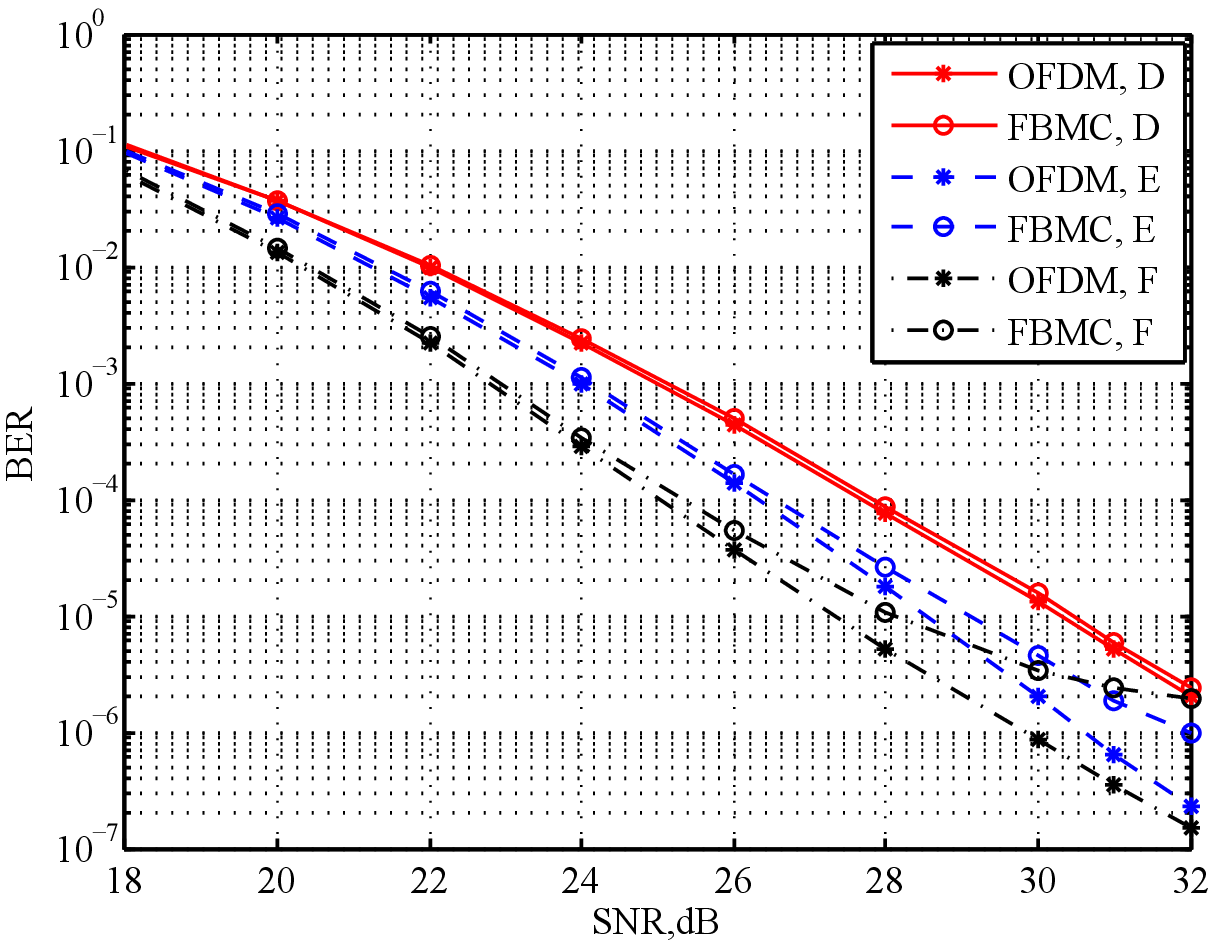}
	\caption{BER performance of the proposed SVD-FS-FBMC system with 64-QAM and rate 2/3 coding, channel Model D, E, and F, $N_{\rm{iter}}=3$.}
	\label{fig:DEF_64QAM_coded}
\end{figure}

\begin{figure}
	\centering
	\includegraphics[scale=0.6]{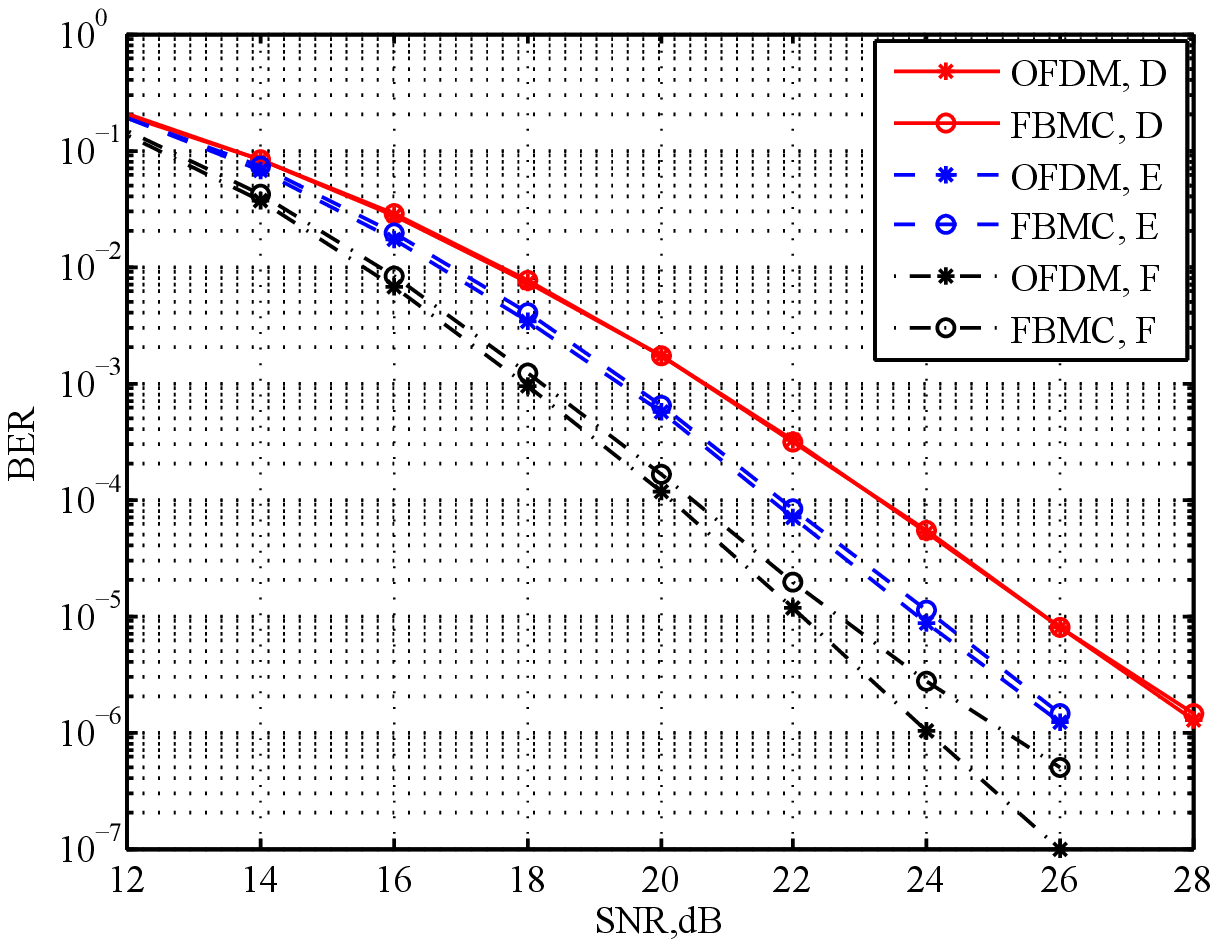}
	\caption{BER performance of the proposed SVD-FS-FBMC system with 16-QAM and rate 2/3 coding, channel Model D, E, and F, $N_{\rm{iter}}=3$.}
	\label{fig:DEF_16QAM_coded}
\end{figure}

Performance of the proposed scheme is also evaluated under channels of more frequency selectivity, i.e., Channel Model E and F, with $2/3$ coding, for $64$-QAM and $16$-QAM modulation, respectively in Fig. \ref{fig:DEF_64QAM_coded} and Fig. \ref{fig:DEF_16QAM_coded}. As the channel selectivity increases, error floor is observed for FBMC systems. Employing lower-order modulation reduces the performance gap between the proposed SVD-FS-FBMC and SVD-OFDM, however does not stop it from growing at high SNRs under the Channel Model F.

\begin{figure}
	\centering
	\includegraphics[scale=0.6]{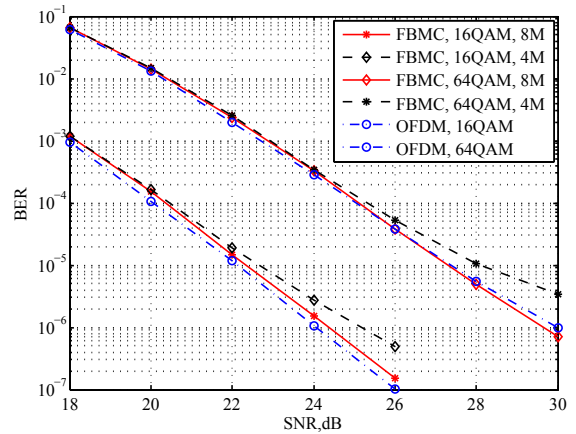}
	\caption{BER performance of the proposed SVD-FS-FBMC systems with 16 and 64-QAM, rate 2/3 coding, FFT size $4M$ and $8M$, $N_{\rm iter}=3$, Channel Model F.}
	\label{fig:comparison_8M_4M}
\end{figure}

In the simulation of Fig. \ref{fig:comparison_8M_4M}, we increase the FFT size to $8M$ and test the proposed SVD-FS-FBMC under channel Model F. Increasing the FFT size from $4M$ to $8M$ doubles the number of frequency tones of the FS-FBMC receiver, and therefore gives even finer beamforming, at the cost of doubled complexity. As observed from Fig. \ref{fig:comparison_8M_4M}, performance of the proposed SVD-FS-FBMC is improved at high SNRs and close to that of SVD-OFDM.

It should be noted that the BER plots above is with respect to SNR. If we use $E_b/N_0$ instead of SNR for x-axis, about 1 dB gain of the proposed SVD-FS-FBMC will be observed over SVD-OFDM, because that the SVD-FS-FBMC does not pay for the energy overhead to transmit the 25\% cyclic prefix as the OFDM in IEEE 802.11n does.

\subsection{Computational complexity of the proposed smoothing methods}\label{simulation_complexity}
In this subsection, we examine the number of iterations required for the smoothing of orthogonal iteration and then give a complexity comparison between the phase factor optimization and orthogonal iteration.

\begin{figure}
	\centering
	\includegraphics[scale=0.6]{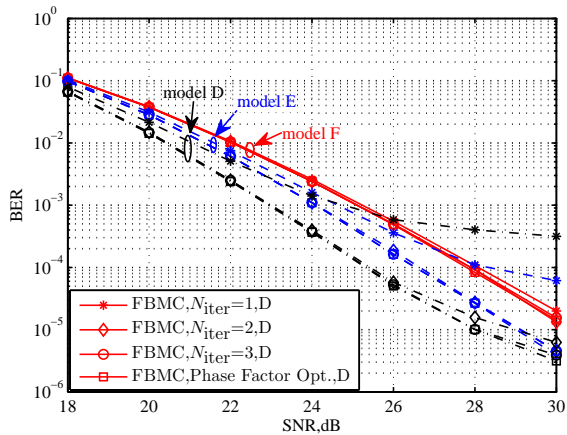}
	\caption{BER performance of the proposed SVD-FS-FBMC system with 64-QAM and rate 2/3 coding, FFT size $4M$, Channel Model F, $N_{\rm iter}=1,2,3$ and $10$.}
	\label{fig:F_iter_phase}
\end{figure}

The BER performance of the proposed SVD-FS-FBMC system with orthogonal iteration is presented for different number of iterations in Fig. \ref{fig:F_iter_phase}, with $2/3$ coding, $64$-QAM modulation, various channel models. Performance of phase factor optimization is also presented for comparison. As observed from Fig. \ref{fig:F_iter_phase}, $N_{\rm{iter}}=3$ is adequate for the orthogonal iteration method to achieve a similar performance as the phase factor optimization under Channel Model D, E and F. For channel model with less frequency selectivity, such as Channel Model D and E, the number of iterations could be further reduced.

\begin{table*}[!t]
	\caption{Complexity of Operations in Smoothed SVD with Orthogonal Iteration.} \label{complexity}
	\centering
	\begin{tabular}{|c|c|c|c|}
		\hline
		Operations & Complexity (FLOPS)\\ \hline
		${{\bf{A}}_k} = {\bf{H}}_k^{\rm{H}}{{\bf{H}}_k}$ & $N_t^2{N_r} + {N_t}{N_r} - \frac{1}{2}N_t^2 - \frac{1}{2}{N_t}$ \\ \hline
		${{\bf{B}}_k} = {{\bf{A}}_k}{{\bf{V}}_k}$, $N_{\rm{iter}}$ times & $(2N_t^3 - N_t^2)N_{\rm{iter}}$ \\ \hline
		QR decomposition: ${{\bf{B}}_k} = {{\bf{V}}_k}{{\bf{R}}_k}$, $N_{\rm{iter}}$ times  & $(\frac{4}{3}N_t^3)N_{\rm{iter}}$  \\ \hline
		Calculation of ${{\bf{U}}_k}$ from ${{\bf{H}}_k} = {\bf{U}}_k{\bf{D}}_k{\bf{V}}_k^{\rm{H}}$
   & $2N_t^2{N_r}$    \\ \hline
	\end{tabular}
\end{table*}

We express the computational complexity in terms of the number of floating point operations (FLOPS) \cite{Hunger2007}. Each scalar/complex addition or multiplication is counted as one FLOPS. Table \ref{complexity} shows the complexity of operations in smoothed SVD with orthogonal iteration \cite{Hunger2007}. Assuming $N_{\rm{iter}}=3$ and ${N_t} = {N_r}$, the total complexity of the orthogonal iteration for each frequency tone is $13N_t^3 - \frac{5}{2}N_t^2 - \frac{1}{2}{N_t}$ FLOPS. Omitting the small order terms, the complexity is $13N_t^3$ FLOPS for each frequency tone. For the phase factor optimization method, the major complexity of each frequency tone is the direct computation of SVD from the channel matrix ${{\bf{H}}_k}$, which is about $4N_t^2N_r+8N_tN_r^2+9N_r^3$ FLOPS as given in \cite{Horn2012}. Assuming ${N_t} = {N_r}$, the complexity of the phase factor optimization is $21N_t^3$ FLOPS for each frequency tone, which is higher than that of the orthogonal iteration with three iterations.

%In this paragraph, we will analysis the complexity between the orthogonal iteration method, where $N_{\rm{iter}}=3$, and the phase factor optimization method. We express the computation in terms of the number of floating point operations (FLOPS) \cite{Hunger2007}. Each scalar/complex addition or multiplication is counted as one FLOPS. Table \ref{complexity} shows the complexity of the operations in the orthogonal iteration method. It is obtained that the total complexity of the orthogonal iteration is $13N_t^3 - \frac{5}{2}N_t^2 - \frac{1}{2}{N_t}$ FLOPS, where ${N_t} = {N_r}$. Omitting the small order terms, the complexity of the orthogonal iteration is $13N_t^3$ FLOPS, which is lower than $4N_t^2N_r+8N_tN_r^2+9N_r^3$ FLOPS (that is $21N_t^3$ FLOPS, if ${N_t} = {N_r}$) in classical SVD \cite{Horn2012}. From the analysis above, it is evident that the complexity of the orthogonal iteration method ($N_{\rm{iter}}=3$) is far below classical SVD. The phase factor optimization changes the phase of the beamforming matrices generated by standard SVD. Therefore, the complexity of the phase factor optimization is higher than standard SVD. Thus, the complexity of the orthogonal iteration is lower than the phase factor optimization method.

%\begin{figure}
%	\centering
%	\includegraphics[scale=0.6]{DEF_64QAM_uncoded.eps}
%	\caption{BER performance of the SVD-FS-FBMC systems with 64-QAM and no coding, in channel model D, E, and F.}
%	\label{fig:DEF_64QAM_uncoded}
%\end{figure}

%\begin{figure}
%	\centering
%	\includegraphics[scale=0.6]{DEF_64QAM_coded_move.eps}
%	\caption{BER performance of the SVD-FS-FBMC systems with 64-QAM and rate 2/3 coding, in channel model D, E, and F.}
%	\label{fig:DEF_64QAM_coded_move}
%\end{figure}
%
%\begin{figure}
%	\centering
%	\includegraphics[scale=0.6]{DEF_16QAM_coded_move.eps}
%	\caption{BER performance of the SVD-FS-FBMC systems with 16-QAM and rate 2/3 coding, in channel model D, E, and F.}
%	\label{fig:DEF_16QAM_coded_move}
%\end{figure}

\section{Conclusions}
%In this paper, we compare the proposed scheme with OFDM in both error performance and complexity. Simulation results have demonstrated that both coded and uncoded error performance of MIMO-OFDM that are close to its OFDM counterpart. This is achieved by effectively combining two techniques: finer granularity and smoothing.

This paper proposed a scheme and a couple of methods to combine SVD beamforming and FBMC/OQAM. Simulation results show that the proposed SVD-FS-FBMC system shares close BER performance with its OFDM counterpart under the IEEE 802.11n Channel Models. The excellent performance comes from two aspects that greatly improve the smoothness of beamformers: i) beamforming with finer granularity in frequency domain; ii) smoothing the beamformers from tone to tone.

%The proposed finer and smoothed beamforming architecture may also serve as a paradigm for the combinations of FBMC/OQAM and other MIMO techniques.

Although the orthogonal iteration reduces the complexity of SVD decomposition to certain level, it is still quite a computational burden when the number of antennas and streams is large. In the future, lower-complexity beamforming schemes and tradeoff between error performance and computational complexity are to be studied.

%Finally, it is worth mentioning that, although this paper employs the FS-FBMC structure to implement the finer granularity beamforming, the proposed scheme could also be adapted to the polyphase network FBMC (PPN-FBMC) structure \cite{Mestre2016,Harris2003,Bellanger2010}, due to the equivalence of these two structures. A design with PPN-FBMC could enjoy considerable reduction of computational complexity. However, very different methods are to be proposed to provide beamformer smoothing with PPN-FBMC, therefore we also leave it for future study.

\appendix
\renewcommand{\appendixname}{Appendix~\Alph{section}}
\section{APPENDIX}
In this appendix, we prove (\ref{eq final}) under the nearly-flat assumptions on ${{\bf{H}}_k}$, ${{\bf{V}}_k}$, ${{\bf{E}}_k}$, and ${{\bf{U}}_k}$. Let (\ref{eq final_pre}) be rewritten as
\begin{equation}\label{eq app_1}
\begin{split}
&{\widetilde{\bf{a}}_{m_0,n_0}} \\
=~&{\bf{c}}_{m_0,n_0} +\sum\limits_{(m,n) \ne (m_0,n_0)}{{\bf{c}}_{m,n} } \\
&+{ {\sum\limits_{k = 0}^{KM - 1} {G_{m_0,n_0}^{(n_0)\,*}(k){{\bf{E}}_k}{{\bf{U}}_k^{\textrm H}}{\bf{n}}^{(n_0)}(k)} } },
\end{split}
\end{equation}
where ${\bf{c}}_{m,n}$ represents the contribution of ${\bf{a}}_{m,n}$ to ${\widetilde{\bf{a}}_{m_0,n_0}}$. Then, our goal in this appendix is to prove ${\bf{c}}_{m,n}\approx {\bf{a}}_{m,n}{\zeta _{m,n}^{m_0,n_0}}$ so that (\ref{eq final}) holds.

%First, let us consider ${\bf{c}}_{m,n_0}$. Let ${\widetilde{\bf{b}}}_{m,n_0}^{(n_0)}(k)$ denote the part of ${\widetilde{\bf{b}}}_{n_0}^{(n_0)}(k)$ that is contributed by ${\bf{a}}_{m,n_0}$.
%Then, ${\bf{c}}_{m,n_0}$ is obtained as
%\begin{equation}\label{eq app_3}
%{\bf{c}}_{m,n_0}= { {\sum\limits_{k = 0}^{KM - 1} {G_{m_0,n_0}^{(n_0)\,*}(k){{\bf{E}}_k}{{\bf{U}}_k^{\textrm H}}{\widetilde{\bf{b}}}_{m,n_0}^{(n_0)}(k)} } }\\.
%\end{equation}
%Due to the assumption that ${\bf{H}}_k$ is nearly flat across the subcarrier band, the channel delay spread is much less than the FFT size $KM$, thus we have the following approximation
%\begin{equation}\label{eq app_2}
%{\widetilde{\bf{b}}}_{m,n_0}^{(n_0)}(k)\approx{\bf{H}}_k {\bf{V}}_k {\bf{a}}_{m,n_0} G_{m,n_0}^{(n_0)}(k).
%\end{equation}
%Using the relation that ${{\bf{U}}_k^{\textrm H}}{{\bf{H}}_k}{{\bf{V}}_k}= {{\bf{D}}_k}$, ${\bf{E}}_k{\bf{D}}_k$ $={\bf{I}}_L$ (assuming the ZF equalization of singular values), we have
%\begin{equation}\label{eq app_4}
%\begin{split}
%{\bf{c}}_{m,n_0}\approx & { {\sum\limits_{k = 0}^{KM - 1} {G_{m_0,n_0}^{(n_0)\,*}(k){\bf{a}}_{m,n_0} G_{m,n_0}^{(n_0)}(k)  } } }\\
%= &~{\bf{a}}_{m,n_0}{\zeta _{m,n_0}^{m_0,n_0}}.
%\end{split}
%\end{equation}
%Specifically, ${\bf{c}}_{m_0,n_0}\approx{\bf{a}}_{m_0,n_0}$.

Combining (\ref{eq final_pre}) and (\ref{eq app_1}), we have
\begin{equation}\label{eq app_main}
{\bf{c}}_{m,n}= { {\sum\limits_{k = 0}^{KM - 1} {G_{m_0,n_0}^{(n_0)\,*}(k){{\bf{E}}_k}{{\bf{U}}_k^{\textrm H}}{\widetilde{\bf{b}}}_{m,n}^{(n_0)}(k)} } },
\end{equation}
where ${\widetilde{\bf{b}}}_{m,n}^{(n_0)}(k)$ denote the part of ${\widetilde{\bf{b}}}^{(n_0)}(k)$ that is contributed by ${\bf{a}}_{m,n}$.

Due to the assumption that ${{\bf{E}}_k}$ and ${{\bf{U}}_k}$ are nearly flat across the subcarrier band, the term $G_{m_0,n_0}^{(n_0)\,*}(k){{\bf{E}}_k}{{\bf{U}}_k^{\textrm H}}$ in (\ref{eq app_main}) can be approximated for $k$ around $Km_0$ as
\begin{equation}\label{eq app_6}
G_{m_0,n_0}^{(n_0)\,*}(k){{\bf{E}}_k}{{\bf{U}}_k^{\textrm H}}\approx G_{m_0,n_0}^{(n_0)\,*}(k){{\bf{E}}_{Km_0}}{{\bf{U}}_{Km_0}^{\textrm H}},
\end{equation}
which then corresponds to $g^*_{m_0,n_0}(i){{\bf{E}}_{Km_0}}{{\bf{U}}_{Km_0}^{\textrm H}}$ in time domain.

On the other hand, at the transmitter, the beamformed signals on all antennas is given in frequency domain by
\begin{equation}\label{eq app_2}
{\bf{b}}_{m,n}(k)={\bf{V}}_k {\bf{a}}_{m,n} G_{m,n}^{(n)}(k).
\end{equation}
It corresponds to  ${\bf{x}}_{m,n}(i)$ in time domain, which is defined as
\begin{equation}\label{eq def_x_mn}
{\bf{x}}_{m,n}(i) =
\begin{cases}
\sum\limits_{k = 0}^{KM - 1}{{\bf{b}}_{m,n}}(k){e^{{\textrm j}\frac{{2\pi k(i-nM/2)}}{{KM}}}},\\ ~~~~~~~~~~~~\frac{nM}{2} \leq i \leq \frac{nM}{2}+KM-1\\
0,~~~~~~~~~~{\rm else}
\end{cases}.
\end{equation}
Due to the assumption that ${{\bf{V}}_k}$ is nearly flat across the subcarrier band,
\begin{equation}\label{eq app_2}
{\bf{b}}_{m,n}(k)\approx {\bf{V}}_{Km} {\bf{a}}_{m,n} G_{m,n}^{(n)}(k),
\end{equation}
and it corresponds to
\begin{equation}\label{eq app_2}
{\bf{x}}_{m,n}(i)\approx {\bf{V}}_{Km} {\bf{a}}_{m,n} g_{m,n}(i).
\end{equation}
Due to the assumption that ${{\bf{H}}_k}$ is nearly flat across the subcarrier band, the signals on all receive antennas that are contributed by ${\bf{a}}_{m,n}$ is
\begin{equation}\label{eq app_ymn}
{\bf{y}}_{m,n}(i)\approx {{\bf{H}}_{Km}}{\bf{x}}_{m,n}(i)\approx {{\bf{H}}_{Km}}{\bf{V}}_{Km} {\bf{a}}_{m,n} g_{m,n}(i).
\end{equation}

Taking the note that ${\widetilde{\bf{b}}}_{m,n}^{(n_0)}(k)$ is the FFT of the part of ${\bf{y}}_{m,n}(i)$ falling into the $n_0$-th window, and using (\ref{eq app_6}) and (\ref{eq app_ymn}), (\ref{eq app_main}) is rewritten in time domain as
\begin{equation}\label{eq app_gg}
\begin{split}
{\bf{c}}_{m,n}= & { {\sum\limits_{k = 0}^{KM - 1} {G_{m_0,n_0}^{(n_0)\,*}(k){{\bf{E}}_k}{{\bf{U}}_k^{\textrm H}}{\widetilde{\bf{b}}}_{m,n}^{(n_0)}(k)} } }\\
\approx & \sum\limits_{i = \frac{n_0 M}{2}}^{\frac{n_0 M}{2}+KM-1}g^*_{m_0,n_0}(i){{\bf{E}}_{Km_0}}{{\bf{U}}_{Km_0}^{\textrm H}}  {\bf{y}}_{m,n}(i)\\
= & \sum\limits_{i = -\infty}^{+\infty}g^*_{m_0,n_0}(i){{\bf{E}}_{Km_0}}{{\bf{U}}_{Km_0}^{\textrm H}}  {\bf{y}}_{m,n}(i)\\
\approx & \sum\limits_{i = -\infty}^{+\infty}g^*_{m_0,n_0}(i){{\bf{E}}_{Km_0}}{{\bf{U}}_{Km_0}^{\textrm H}}  {{\bf{H}}_{Km}}{\bf{V}}_{Km} {\bf{a}}_{m,n} g_{m,n}(i).
\end{split}
\end{equation}
Clearly, ${\bf{c}}_{m,n}$ is non-zero only when $g^*_{m_0,n_0}(i)$ and $g_{m,n}(i)$ overlap in frequency, which means that frequency tone $Km_0$ and $Km$ are within the width of one subcarrier band. Taking use of the nearly-flat assumption, we replace ${{\bf{E}}_{Km_0}}$ and ${{\bf{U}}_{Km_0}^{\textrm H}}$ in (\ref{eq app_gg}) by ${{\bf{E}}_{Km}}$ and ${{\bf{U}}_{Km}^{\textrm H}}$, respectively. Finally, using the relations that ${{\bf{U}}_k^{\textrm H}}{{\bf{H}}_k}{{\bf{V}}_k}= {{\bf{D}}_k}$ and ${\bf{E}}_k{\bf{D}}_k$ $={\bf{I}}_L$ (assuming the ZF equalization of singular values), we arrive at
\begin{equation}\label{eq app_5}
{\bf{c}}_{m,n}\approx {\bf{a}}_{m,n}{\zeta _{m,n}^{m_0,n_0}}.
\end{equation}
Specifically, ${\bf{c}}_{m_0,n_0}\approx{\bf{a}}_{m_0,n_0}$.

\end{document}